\documentclass[]{spie}  

\usepackage{amsmath,amsfonts,amssymb}
\usepackage{graphicx}
\usepackage{caption}
\usepackage{float}
\usepackage[colorlinks=true, allcolors=blue]{hyperref}

\title{The Wide-field Spectroscopic Telescope (WST): design trade-offs for the low-resolution multi-object spectrograph instrument}

\author[a]{D. Buffat}
\author[b]{W. Saunders}
\author[a]{K. Dohlen}
\author[a]{L. Tresse}
\author[c]{D. Lee}
\author[d]{R. Bacon}
\author[e]{A. Bianco}
\author[f]{L. Fréour}
\author[g]{R. de Jong}
\author[h]{E. Burtin}
\author[i]{O. Iwert}
\author[h]{V. Meyer}
\author[i]{E. George}

\affil[a]{Aix-Marseille Université, CNRS, CNES, LAM UMR7326, Marseille, France}
\affil[b]{Australian Astronomical Optics, Macquarie Univ., Sydney, Australia}
\affil[c]{STFC UK Astronomy Technology Centre, Royal Observatory Edinburgh, Blackford Hill,
Edinburgh, EH9 3HJ, United Kingdom}
\affil[d]{Université Claude Bernard Lyon 1, CNRS, Centre de Recherche Astrophysique de Lyon,
UMR5574, Saint-Genis-Laval, France}
\affil[e]{INAF, Osservatorio Astronomico di Brera, Milano, Italy}
\affil[f]{Department of Astrophysics, University of Vienna}
\affil[g]{Leibniz-Institut für Astrophysik Potsdam (AIP), Potsdam, Germany}
\affil[h]{CEA, Paris-Saclay, France}
\affil[i]{European Southern Observatory, Garching, Germany}

\authorinfo{Send correspondence to dimitri.buffat@lam.fr}

\begin{document} 
\maketitle

\begin{abstract}
The Wide-field Spectroscopic Telescope\footnote{\url{https://wstelescope.eu}} (WST) is a planned 12-meter-class dedicated spectroscopic facility designed to address key scientific challenges through large spectroscopic surveys. This paper presents design and performance trade-offs for the Low-Resolution Multi-Object Spectrograph (MOS-LR) instrument. With a multiplex of 30,000 covering a field of view of 3.1 square degrees, this instrument will provide unprecedented survey efficiency, an order of magnitude beyond those of current facilities. Covering the 370 to 930 nm range at a resolving power of 3,000 with a sky-projected fiber diameter of 1 arcsec, this instrument faces extreme challenges in design, manufacturing, and maintenance. We present a systematic approach to trading off optical and mechanical design options, taking into account constraints such as volume and mass, but also projected availability of detectors and gratings etc. 
\end{abstract}


\section{INTRODUCTION}
\label{sec:intro}  

\subsection{General context}

The advent of large surveys (e.g., LSST, Euclid, Roman, ELT, SKAO, CTAO) has highlighted the need for a large spectroscopic survey. Spectroscopic surveys already exist using Multi-Object Spectrographs (MOS), such as Mayall/DESI [\citenum{Perruchot2020}], VISTA/4MOST [\citenum{Caillier2018}], WHT/WEAVE [\citenum{Rogers2014}], VLT/MOONS [\citenum{Oliva2016}], or Subaru/PFS [\citenum{Vivs2012}], which possess a multiplex ranging from a few hundred to a few thousand fibres. In parallel, there are also surveys using Integral Field Spectrographs (IFS) such as VLT/MUSE [\citenum{Cai2025}], or ELT/HARMONI [\citenum{Loupias2020}]. WST is a project that combines these two instrumental concepts. Composed of both high and low resolution MOS (referred to as MOS-HR and MOS-LR respectively). Thus simultaneously and independently producing three surveys allowing a wide variety of objects to be observed and covering a large number of scientific cases [\citenum{WST_WhitePaper2024}].

\subsection{Current landscape}

To address this need, the WST project aims to combine the strengths of MOS and IFS instruments. Below, we compare WST's capabilities with existing facilities. In the context of multi-messenger astrophysics and transient phenomena, a large étendue is essential to observe a maximum number of objects. The étendue of a telescope is defined as $G = A \times \Omega$. Where A is the area of the primary mirror, and $\Omega$ the solid angle of the field of view. In Tables \ref{tab:mos_etendue} and \ref{tab:ifs_etendue}, WST does not have the telescope with the largest M1. However, it is the telescope with the largest étendue for both a MOS and an IFS, making it a true spectrum factory.

\begin{table}[h]
    \centering
    \begin{tabular}{l|l|l|l}
    Telescope & Primary mirror diameter (m) & FoV (deg$^2$) & Etendue (m$^2 \cdot$deg$^2$) \\
    \hline
    WST/MOS   & 12 & 3.1 & 351 \\
    Rubin/LSST & 8.4 & 9.6 & 319 \\
    Subaru/HSC & 8.2 & 1.8 & 50 \\
    ELT/MOSAIC & 39 & 0.011 & 13 \\
    VLT/MOONS & 8.2 & 0.14 & 7
    \end{tabular}
    \caption{Comparison of WST MOS etendue with other facilities.}
    \label{tab:mos_etendue}
\end{table}

\begin{table}[h]
    \centering
    \begin{tabular}{l|l|l|l}
    Telescope & Primary mirror diameter (m) & FoV & Etendue (m$^2 \cdot$arcmin$^2$) \\
    \hline
    WST/IFS & 12 & 3x3 arcmin$^2$ & 900 \\
    VLT/MUSE & 8.2 & 1x1 arcmin$^2$ & 50 \\
    Keck/KCWI & 10 & 20x30 arcsec$^2$ & 13 \\
    ELT/HARMONI & 39 & 4x3 arcsec$^2$ & 4
    \end{tabular}
    \caption{Comparison of WST IFS field of view and etendue with other IFS.}
    \label{tab:ifs_etendue}
\end{table}

If we need a factory, it is for Rubin/LSST which is a 10-year photometric survey capable of scanning the southern sky in three nights. Having a large spectroscopic survey with WST is in line with the current multi-messenger astrophysics need. In total, WST will have 32,000 fibres of which 30,000 for the MOS-LR. Table \ref{tab:mos_multiplex} compares the multiplex of different low-resolution MOS. Apart from MUST which is currently under construction, we are crossing a non-negligible gap compared to current surveys such as DESI.

\begin{table}[h]
    \centering
    \begin{tabular}{l|l|l}
    Instrument & MOS multiplex & Detector size (number of pixels x pixel size)\\
    \hline
    WST/MOS-LR & 30,000 & TBD \\
    MUST & 20,000 & 4k x 15 $\mu m$ = 6 cm\\
    Mayall/DESI & 5,000 &  4k x 15 $\mu m$ = 6 cm\\
    Subaru/PFS & 2,400 &  4k x 15 $\mu m$ = 6 cm\\
    VISTA/4MOST-LRS & 2,400 & 6k x 15 $\mu m$ = 9 cm\\
    VLT/MOONS & 1,000 & 4k x 15 $\mu m$ = 6 cm
    \end{tabular}
    \caption{Comparison of WST MOS-LR with other MOS facilities.}
    \label{tab:mos_multiplex}
\end{table}

\section{INSTRUMENTAL REQUIREMENTS}
\label{sec:requir}

\subsection{Top-level requirements}

As explained in Section \ref{sec:intro}, we are considering a multiplex of 30,000 fibres. Regarding the spectral coverage, the work presented here was conducted over the 370–980 nm range. In Section \ref{sec:evol}, we discuss the impacts on the selected design.

\begin{figure}[h]
    \centering
    \includegraphics[width=0.5\linewidth]{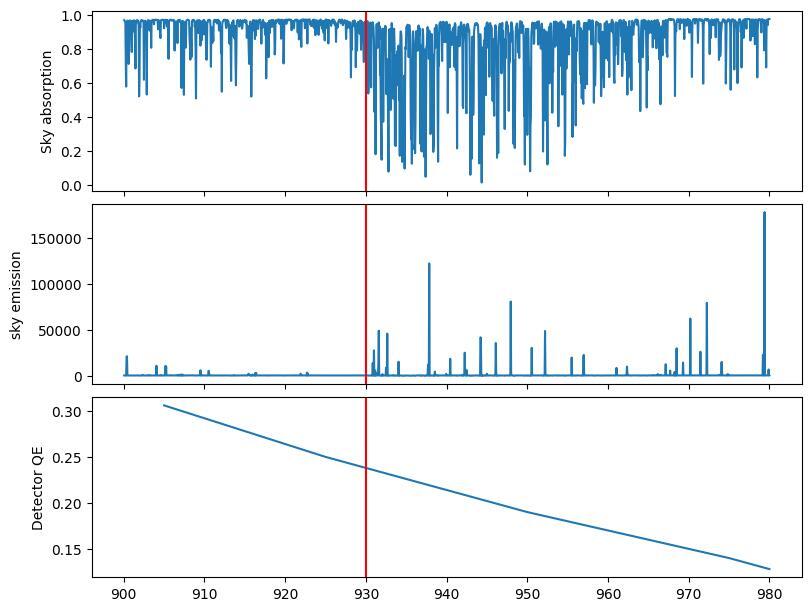}
    \caption{NIR issue.}
    \label{fig:red_lim}
\end{figure}

In the meantime, the reason we changed the spectral coverage is due to the current limitations of CMOS detectors. When combined with sky absorption and emission (Figure \ref{fig:red_lim}), the flux received by the detector would be prohibitively low. The selected design will therefore be re-optimised for this new spectral range to constitute the reference design. However, by 2040, CMOS detector technology will have had time to mature, allowing us to push further into the NIR.

Another requirement is the spectral resolution with $R > 3000$ everywhere. But in the red arm we will seek a higher resolution with $R > 4000$.

For the fibres, we are working with a core diameter of 200 µm, which allows for a projected fibre size of 1 arcsec.

\subsection{Self-imposed requirements}

We imposed an image quality requirement, setting an RMS spot radius $<d/6$, the value achieved by DESI [\citenum{Perruchot2020}], as the upper limit. This criterion was adopted as a design requirement following the example of the DESI spectrograph, which has demonstrated this level of optical performance across its full field and spectral range. Using this value as a reference provides a realistic and proven benchmark, ensuring that the optical design does not become the limiting factor for spectral resolution, while remaining consistent with what has already been achieved in a comparable fiber-fed multi-object spectrograph.

\section{OPTICAL DESIGN CONCEPTS}
\label{sec:design}

\subsection{First order study}

To understand the choice of the explored designs, it is important to carry out a first-order study. This study incorporates several approximations, including a flat slit and constant resolution (i.e. $R_{\text{min}}$ and $R_{\text{max}}$ identical in each arm of the spectrograph).

\begin{equation}
    L_{\text{slit}} = N_{\text{fib}} d_{\text{fib}} \tau
    \label{eq:slit}
\end{equation}

We begin with the slit size defined by Equation (\ref{eq:slit}). Assuming a flat slit, the slit length depends on the number of fibres $N_{\text{fib}}$, the fibre core diameter $d_{\text{fib}}$, and the fibre spacing ratio $\tau$.

\begin{equation}
    \tau = \frac{d_{\text{fib}} + l_{\text{gap}}}{d_{\text{fib}}}
    \label{eq:tau}
\end{equation}

We define $\tau$ with Equation (\ref{eq:tau}) as being the sum of the fibre diameter and the centre-to-centre spacing of two neighbouring fibres, divided by the fibre diameter.

\begin{equation}
    L_{\text{slitDet}} = L_{\text{slit}} \frac{F_{\text{cam}}}{F_{\text{coll}}}
    \label{eq:slit_det}
\end{equation}

The projection of the slit onto the detector is then given by Equation (\ref{eq:slit_det}). Where $F_{\text{cam}}$ and $F_{\text{coll}}$ are respectively the f-number of the camera and the collimator.

\begin{equation}
    F_{\text{coll}} = \frac{F_{\text{tel}}}{1+ F_{\text{tel}} \cdot w_{\text{FRD}}}
    \label{eq:fcoll}
\end{equation}

Equation (\ref{eq:fcoll}) allows us to express the collimator f-number as a function of the telescope f-number and $w_{\text{FRD}}$. The $w_{\text{FRD}}$ is the Focal Ratio Degradation (FRD) coefficient generated by the fibre with respect to $F_{\text{tel}}$ in degrees.

\begin{equation}
    y_{\text{Det}} = N_{\text{fib}} d_{\text{fib}} \tau \left ( \frac{F_{\text{cam}}}{F_{\text{tel}}} + F_{\text{cam}} \cdot w_{\text{FRD}} \right )
    \label{eq:y_det}
\end{equation}

If we make the assumption that the detector size along the $y$ axis is equal to the projected slit size, we obtain Equation (\ref{eq:y_det}).

\begin{equation}
    x_{\text{Det, }i} = N_{\text{pix, }i} \cdot dx_{\text{pix}}
    \label{eq:x_npix}
\end{equation}

In this case, the $x$ axis corresponds to the spectral axis that we must define in each arm. Working backwards, the detector size along this axis is given by Equation (\ref{eq:x_npix}). Where $N_{\text{pix, }i}$ is the number of illuminated pixels in arm $i$ and $dx_{\text{pix}}$ is the pixel size.

\begin{equation}
    N_{\text{pix, }i} = N_{\text{spec, }i} \cdot k_{\text{pix}}
    \label{eq:n_pix}
\end{equation}

\begin{equation}
    N_{\text{spec, }i} = \frac{\lambda_{\text{max, }i} - \lambda_{\text{min, }i}}{d\lambda_i}
    \label{eq:n_spec}
\end{equation}

\begin{equation}
    k_{\text{pix}} = d_{\text{fib}} \frac{\sqrt{3}}{2} \frac{F_{\text{cam}}}{F_{\text{coll}}} \frac{1}{dx_{\text{pix}}}
    \label{eq:k_pix}
\end{equation}

$N_{\text{pix, }i}$ can be expressed according to Equation (\ref{eq:n_pix}) as a function of the number of spectral pixels $N_{\text{spec, }i}$ in each arm (Equation (\ref{eq:n_spec})) and the pixel sampling $k_{\text{pix}}$ (Equation (\ref{eq:k_pix})). Where the term $d_{\text{fib}} \frac{\sqrt{3}}{2}$ is the Full Width at Half Maximum (FWHM) for a perfect system.

\begin{equation}
    \lambda_{\text{min, }i} = \lambda_{j} \left( 1-\frac{\delta\lambda}{2} \right)
    \label{eq:lam_min}
\end{equation}

\begin{equation}
    \lambda_{\text{max, }i} = \lambda_{j+1} \left(1+\frac{\delta\lambda}{2} \right)
    \label{eq:lam_max}
\end{equation}

Regarding $\lambda_{\text{min, }i}$ and $\lambda_{\text{max, }i}$, $i$ corresponds to the arm in which we are working. We will now use the index $j$ which corresponds to the cut-off wavelengths of each arm. This change of variable allows us to write Equations (\ref{eq:lam_min}) and (\ref{eq:lam_max}) which take into account the dichroic overlap $\delta\lambda$.

\begin{equation}
    \lambda_j = \lambda_{\text{min}}^{\frac{N+1-i}{N}} \lambda_{\text{max}}^{\frac{i-1}{N}}
    \label{eq:lam_j}
\end{equation}

The characteristic wavelengths of the system are the $\lambda_j$ where $j \in [1, N+1]$, where $N$ is the number of arms. This means that $\lambda_1 = \lambda_{\text{min}}$ and $\lambda_{N+1} = \lambda_{\text{max}}$. To express $\lambda_j$, we use Equation (\ref{eq:lam_j}) which can be constructed graphically by taking $R_{\text{min, }i}$ and $R_{\text{max, }i}$ as constants.

\begin{equation}
    \lambda'_{\text{min}} = \frac{\lambda_{\text{min}}}{1-\frac{\delta\lambda}{2}}
    \label{eq:lam_min_p}
\end{equation}

\begin{equation}
    \lambda'_{\text{max}} = \frac{\lambda_{\text{max}}}{1+\frac{\delta\lambda}{2}}
    \label{eq:lam_max_p}
\end{equation}

The problem with this last formula is that an edge effect appears at the extreme wavelengths due to the $\delta\lambda$. It is as if a dichroic were being placed at the spectral extremities of the system. This problem is easily solved with a change of variable: the inverse of the overlap is applied to the extreme wavelengths (Equations (\ref{eq:lam_min_p}) and (\ref{eq:lam_max_p})).

\begin{equation}
    \lambda_j = {\lambda'}_{\text{min}}^{\frac{N+1-i}{N}} {\lambda'}_{\text{max}}^{\frac{i-1}{N}}
    \label{eq:lam_j_p}
\end{equation}

\begin{equation}
    d\lambda_{i} = \frac{\lambda_{\text{min, }i}}{R_{\text{LR}}} = \frac{\lambda_j}{R_{\text{LR}}}
    \label{eq:dlam}
\end{equation}

\begin{equation}
    x_{\text{Det}} = \left ( \frac{\lambda'_{\text{min}}}{\lambda'_{\text{max}}} \right )^\frac{1}{N} \left [ \left ( 1 + \frac{\delta \lambda}{2} \right ) - \left ( \frac{\lambda'_{\text{min}}}{\lambda'_{\text{max}}} \right )^\frac{1}{N} \left ( 1 - \frac{\delta \lambda}{2}\right ) \right ] d_{\text{fib}} \frac{\sqrt{3}}{2} R_{\text{LR}} \left ( \frac{F_{\text{cam}}}{F_{\text{tel}}} + F_{\text{cam}} \cdot w_{\text{FRD}} \right )
    \label{eq:x_det}
\end{equation}

By applying this change of variable in Equation (\ref{eq:lam_j}) to obtain Equation (\ref{eq:lam_j_p}). And by writing Equation (\ref{eq:dlam}). We can finally obtain Equation (\ref{eq:x_det}). It will be noted that the various changes of variables have made it possible to obtain a constant number of spectral pixels in each arm of the spectrograph.

It should be noted that x and y have the same dimensions as the fibre diameter $d_{\text{fib}}$.

\begin{figure}[h]
\begin{minipage}{0.80\textwidth}
    \centering
    \includegraphics[width=\linewidth]{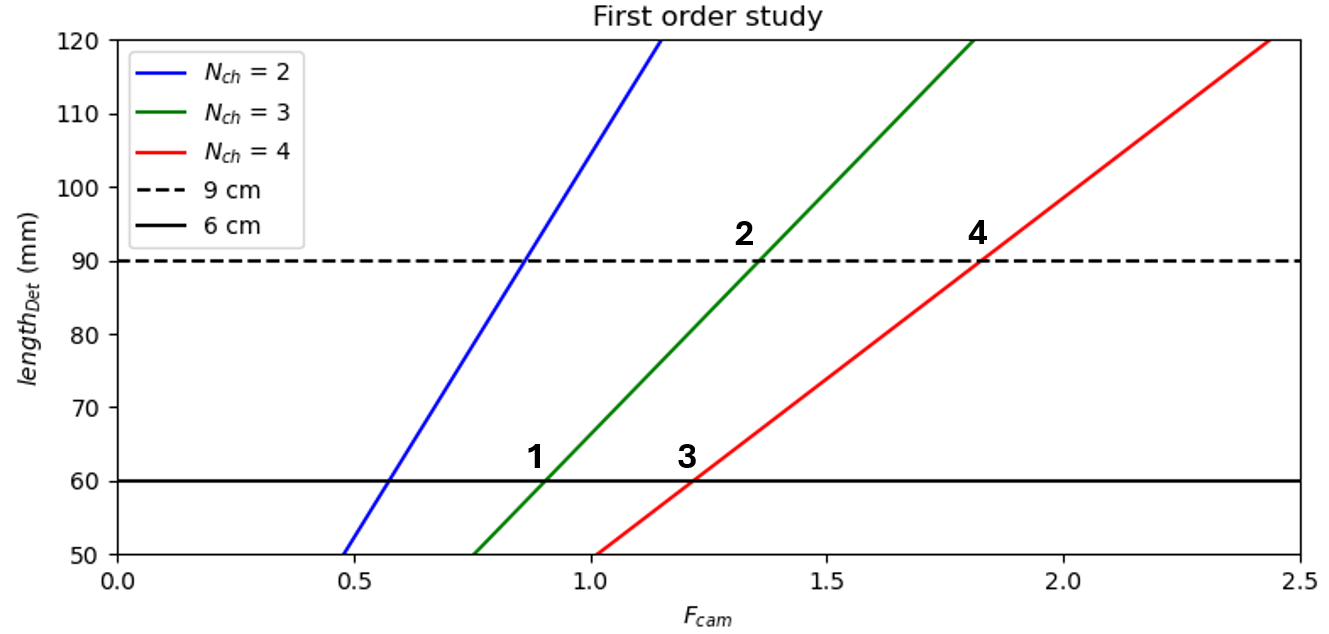}
\end{minipage}
\hfill
\begin{minipage}{0.18\textwidth}
    \centering
    \begin{tabular}{c|l}
        1 & FSS \\
        2 & 3x9 \\
        3 & 4x6 \\
        4 & 4x9
    \end{tabular}
\end{minipage}
\caption{First order study : Detector size as a function of camera f-number for 3-arm and 4-arm configurations, with detector sizes of 6 cm and 9 cm.}
\label{fig:xdet_vs_fcam}
\end{figure}

Finally, we favoured the spectral dimension with $x_{\text{Det}}$ to obtain the detector size as a function of the camera f-number for several concepts with different numbers of arms. This is shown in Figure \ref{fig:xdet_vs_fcam}. We are interested in two detector sizes: 6 cm and 9 cm. As these are the most common sizes in low-resolution spectrographs today (see Table \ref{tab:mos_multiplex}).

According to Figure \ref{fig:xdet_vs_fcam}, it is not of interest to focus on a two-arm design. The f-number being much smaller than 1, this implies complex or free-form optics. We will therefore only consider 4 designs which will be denoted AxB. Where A is the number of arms, and B is the detector size in cm.

The number of arms A is important. The greater this number, the better the efficiency of the diffraction gratings, here Volume Phase Holographic gratings (VPHG) [\citenum{Bianco2026}]. However, the camera f-number, the number of optical elements, and the number of spectrographs all increase. We therefore expect the cost of the MOS-LR to be higher for 4-arm spectrographs. This will be studied in Section \ref{sec:tradeoff}.

The list is given in Figure \ref{fig:xdet_vs_fcam} (right). Note that 3x6 is the name given in the 1st order study. Subsequently it will be referred to as the Folded Solid Schmidt (FSS).

\begin{equation}
    N_{\text{fib}} = \left( \frac{\lambda'_{\text{min}}}{\lambda'_{\text{max}}} \right)^{\frac{1}{N}} \left[ \left( 1 + \frac{\delta\lambda}{2} \right) - \left( \frac{\lambda'_{\text{min}}}{\lambda'_{\text{max}}} \right)^{\frac{1}{N}} \left( 1 - \frac{\delta\lambda}{2} \right) \right] \frac{\sqrt{3}}{2} \frac{ R_{\text{LR}}}{\tau}
    \label{eq:n_fib}
\end{equation}

\begin{equation}
    N_{\text{spectro}} = \frac{\text{Multiplex}}{N_{\text{fib}}}
    \label{eq:n_spectro}
\end{equation}

One final remark concerns the detectors. As the detectors are square, we naturally have $x_{\text{Det}} = y_{\text{Det}}$. From which we can isolate the number of fibres $N_{fib}$ (Equation (\ref{eq:n_fib})) and obtain a number of spectrographs (Equation (\ref{eq:n_spectro})). It will be noted that we have lost the dependence on the f-numbers and the detector size. So, this provides a first approximation of the number of spectrographs for a three or four-arm configuration.

\subsection{Exploration of concepts}

\begin{figure}[h]
    \centering
    \begin{tabular}{c|c|c}
     & 3-arms & 4-arms \\
    6 cm & \includegraphics[width=0.41\textwidth]{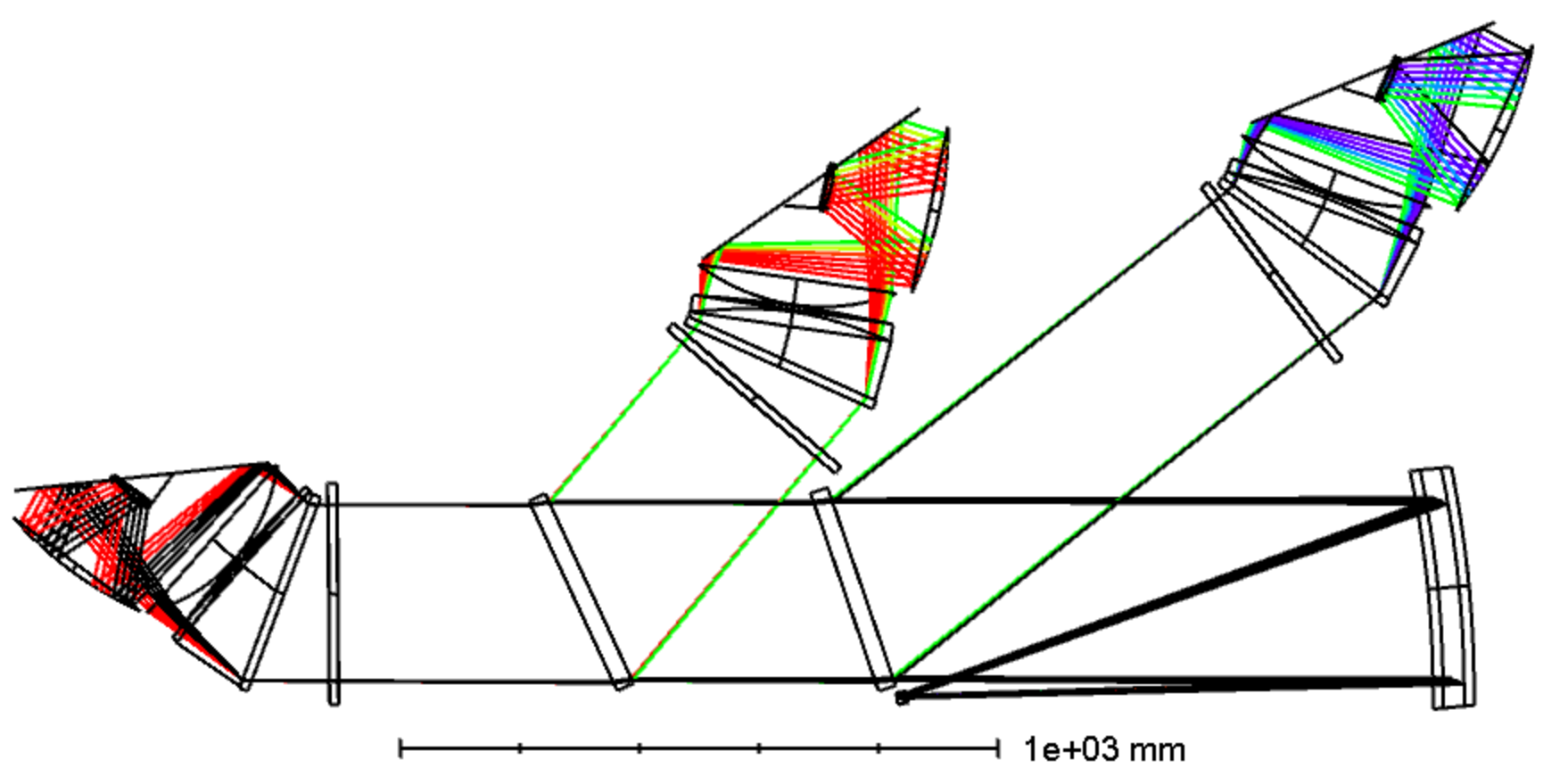} & \includegraphics[width=0.252\textwidth]{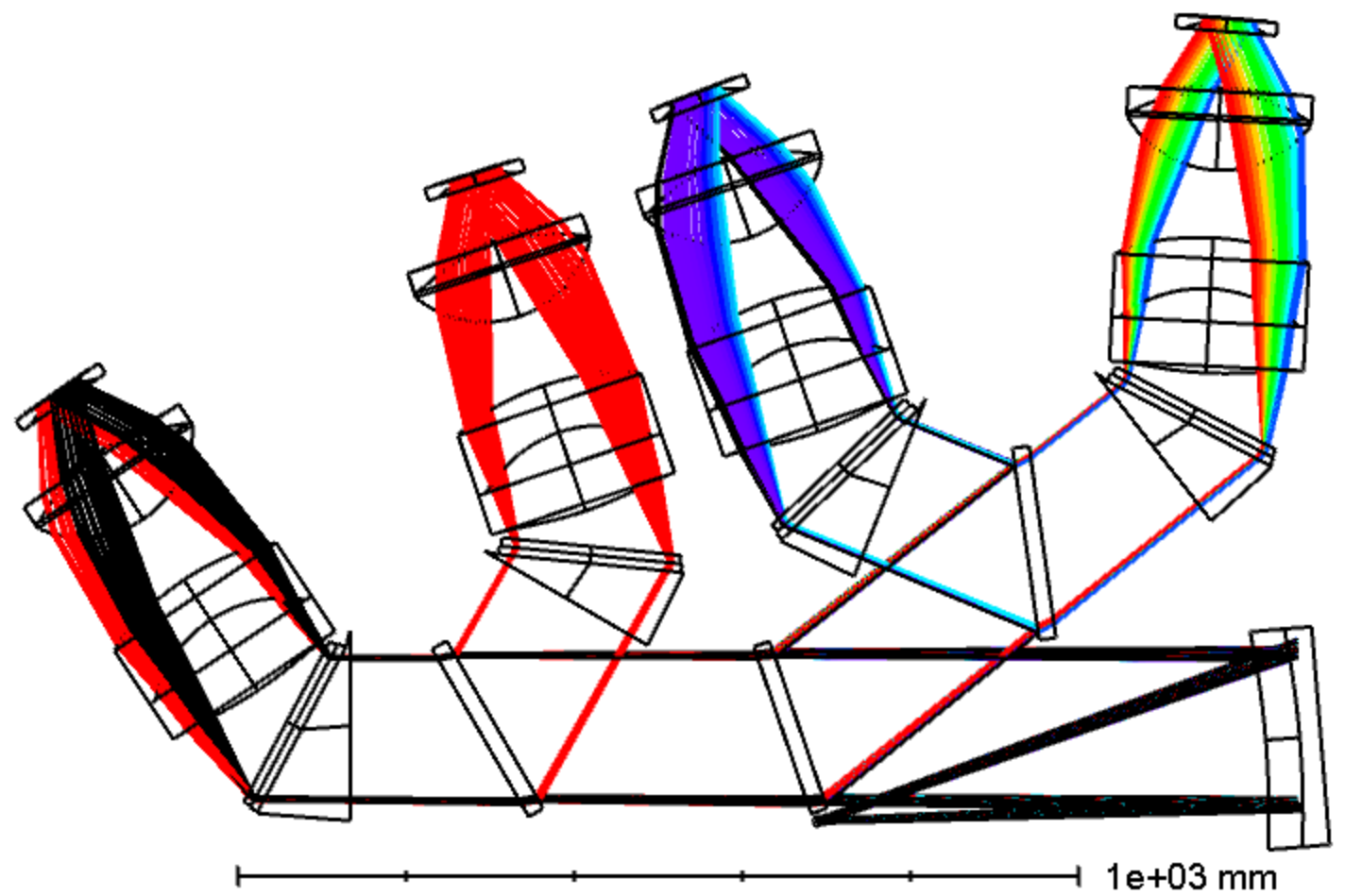} \\
    \hline
    9 cm & \includegraphics[width=0.405\textwidth]{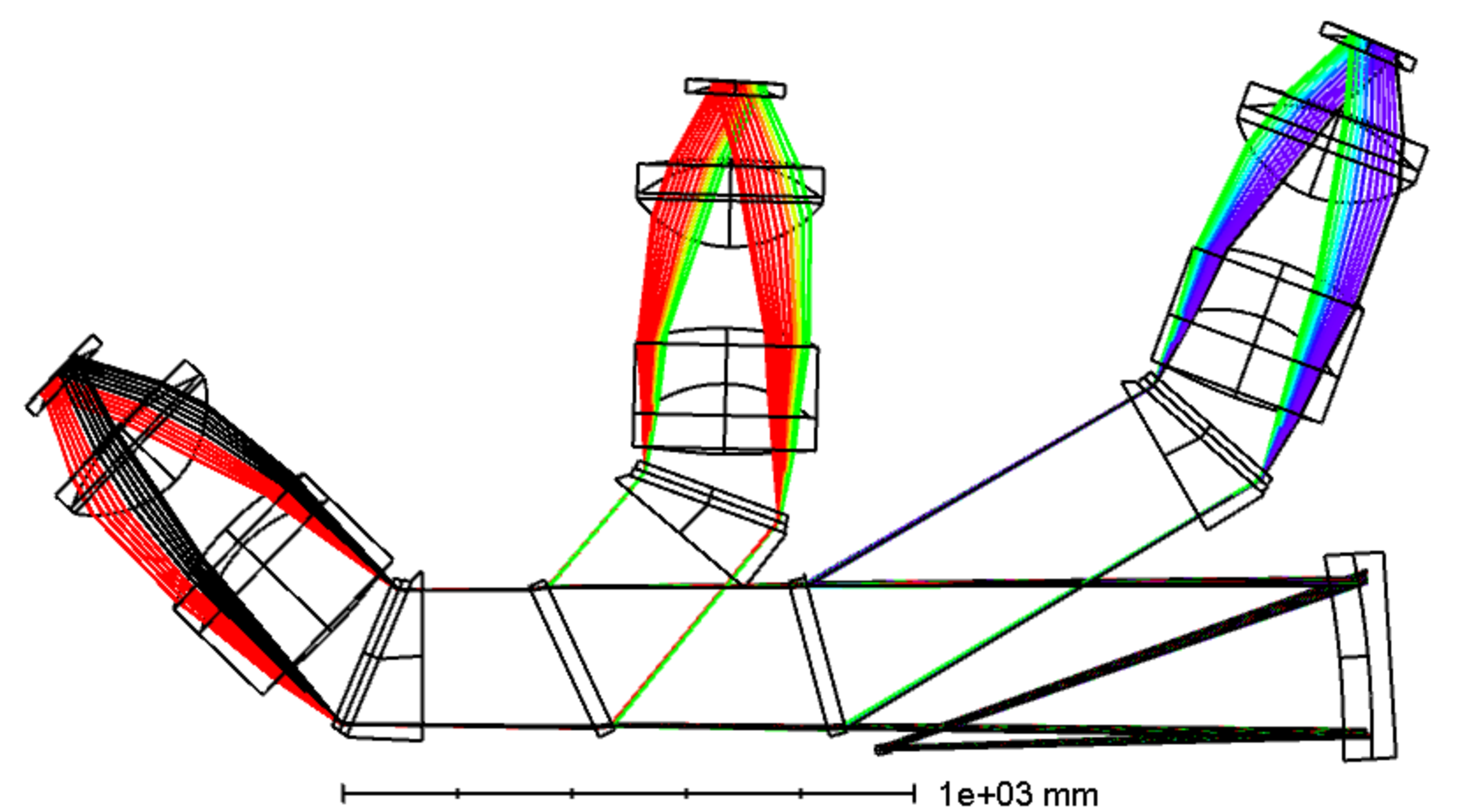} & \includegraphics[width=0.275\textwidth]{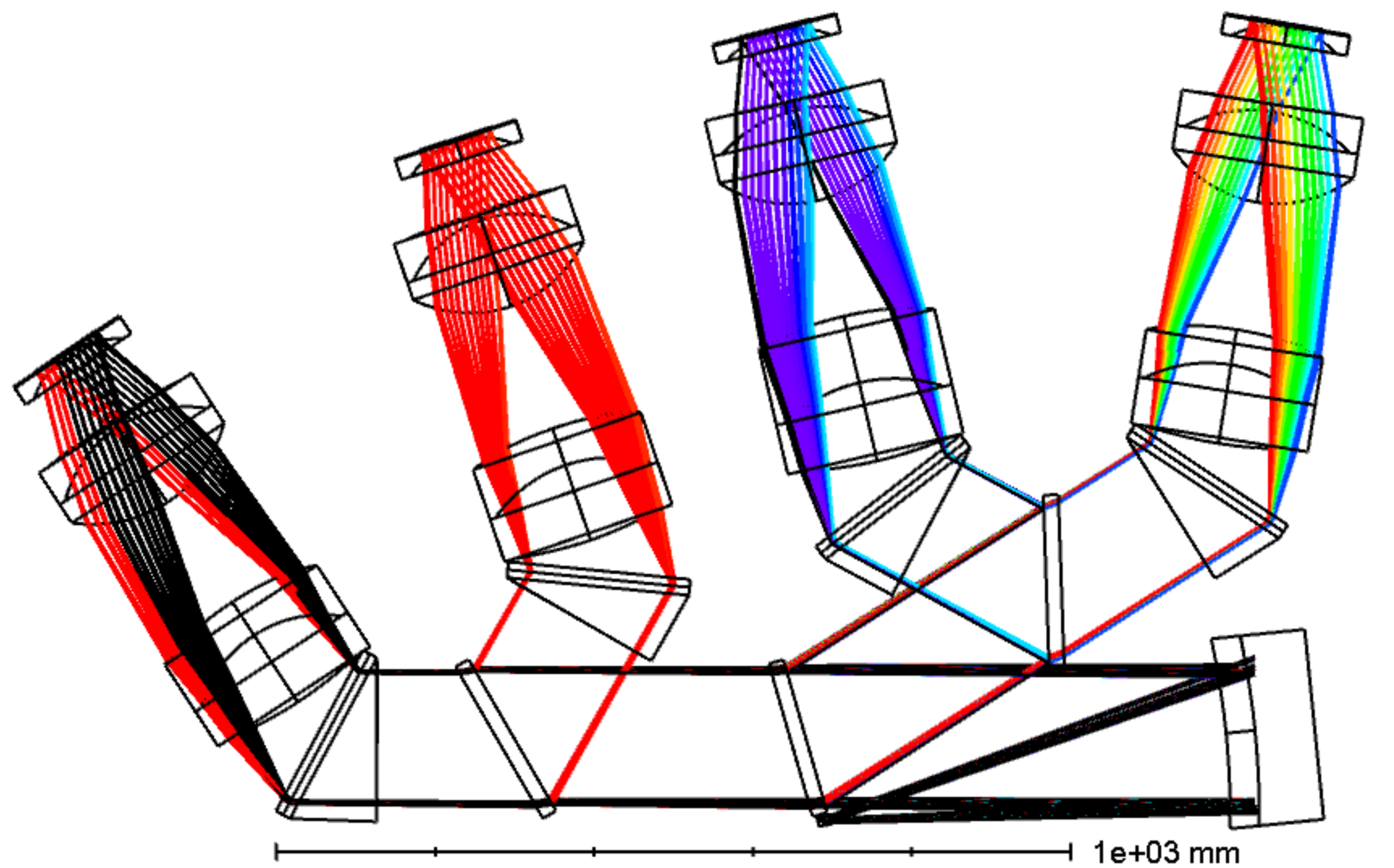}
    \end{tabular}
    \caption{Overall of optical designs concepts at the same scale.}
    \label{fig:ConceptDesignsOverall}
\end{figure}

Figure \ref{fig:ConceptDesignsOverall} presents the optical concepts under study [\citenum{Saunders2026}]. A fifth optical design, a variant of one of the concepts, will be presented at the end of this section.

All the designs produced at this stage meet the requirements. Each of them will be briefly presented in the following subsections.

We can already note some similarities between the different designs: it should be noted that they are all built with an off-axis collimator. That the blue arm is the first to be isolated in order to preserve a maximum of flux. Another interesting point, the slit and the detectors are positioned on opposite faces, which could facilitate access and maintenance on either side of the spectrographs.

Finally, the similarity of the camera f-numbers with the first order study (Table \ref{tab:compare_first_order}, validating our model in the process.

\begin{table}[h]
    \centering
    \begin{tabular}{l|l|l||l|l}
         & \multicolumn{2}{|c||}{\boldmath$F_{cam}$} & \multicolumn{2}{|c}{\boldmath$N_{spectro}$} \\
         & First order & Optical design & First order & Optical design\\
    3x6 / FSS & 0.9 & 0.775 & 43 & 41\\
    3x9 & 1.36 & 1.31 & 43 & 48 \\
    4x6 & 1.22 & 1.17 & 57 & 58 \\
    4x9 & 1.82 & 1.75 & 57 & 58
    \end{tabular}
    \caption{Comparison between the first order study and the optical designs.}
    \label{tab:compare_first_order}
\end{table}

\subsubsection{Folded Solid Schmidt}

Among our designs, the FSS in a 3-arms spectrograph with 6 cm detectors, has a spectral resolution $R > 3100$ everywhere and $R > 5000$ at 860 nm. The spectral coverage and the spectral resolution of each arm is detail in Table \ref{tab:FSS_cov}. 

\begin{table}[h]
    \centering
    \begin{tabular}{l|c|c|c}
         & UB & VR & IZ \\
        \hline
    $\lambda_{min}$ & 370 & 525 & 740 \\
    $\lambda_{max}$ & 540 & 760 & 980 \\
    $R_{min}$ & 3100 & 3100 & 4302 \\
    $R_{max}$ & 4524 & 4488 & 5698
    \end{tabular}
    \caption{Spectral coverage and spectral resolution in each arm for the FSS.}
    \label{tab:FSS_cov}
\end{table}

The FSS is the only one with a catadioptric camera and it is also the fastest camera with f/0.775 and provides an RMS $< d/10$. And it is precisely its camera that can be impressive. Therefore the entrance lens, and the two mirrors that make up the camera are the faces of one single glass block. The detector is then placed at the centre of the flat mirror, with its cryostat which must be fixed to the field lens, located in this glass block which will have been hollowed out beforehand. Our designs were conceived to operate at room temperature.

\begin{figure}[h]
    \centering
    \includegraphics[width=0.5\linewidth]{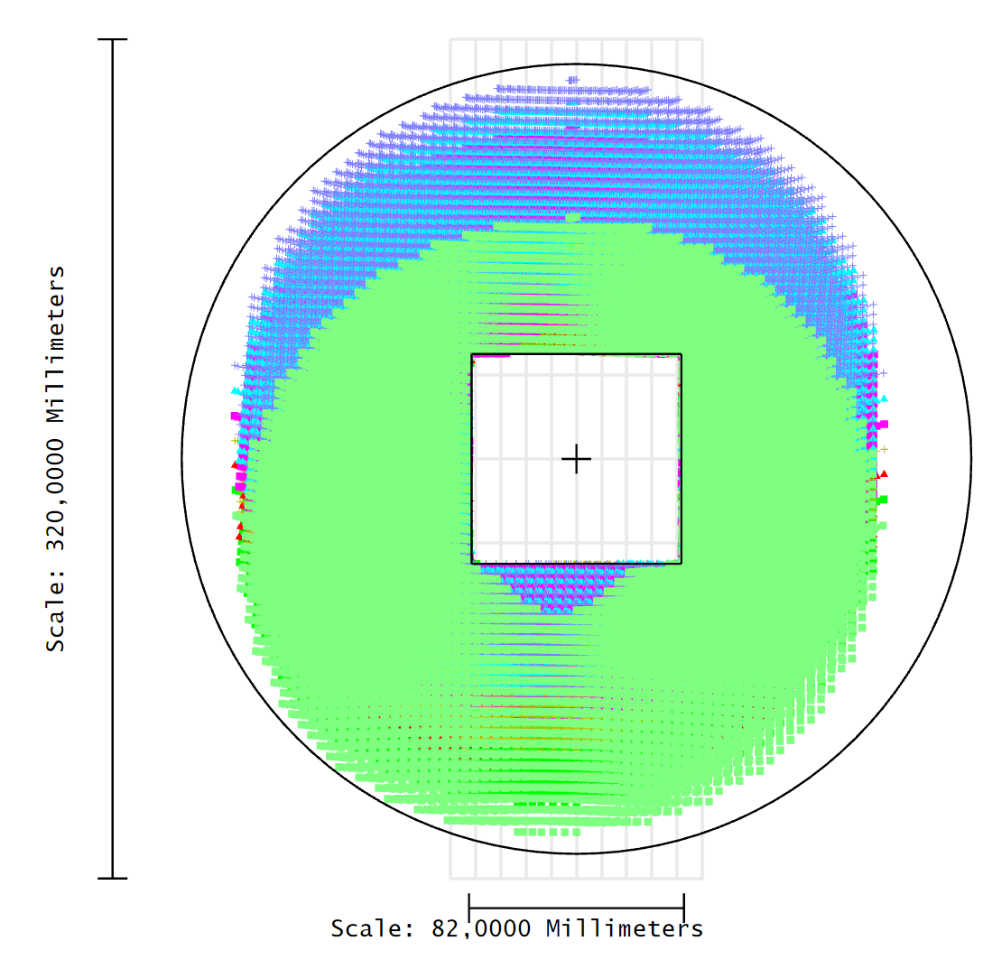}
    \caption{Footprint diagram at the obscuration in the channel VR.}
    \label{fig:fss_tp}
\end{figure}

One of the problems with the detector inside this glass block is stray light. To prevent the centre of the beam from falling onto the detector, we must place an obscuration. This obscuration blocks $\sim 20\%$ of the total flux in each arm (Figure \ref{fig:fss_tp}).

\begin{figure}[h]
    \centering
    \begin{tabular}{ccc}
    \includegraphics[width=0.25\linewidth]{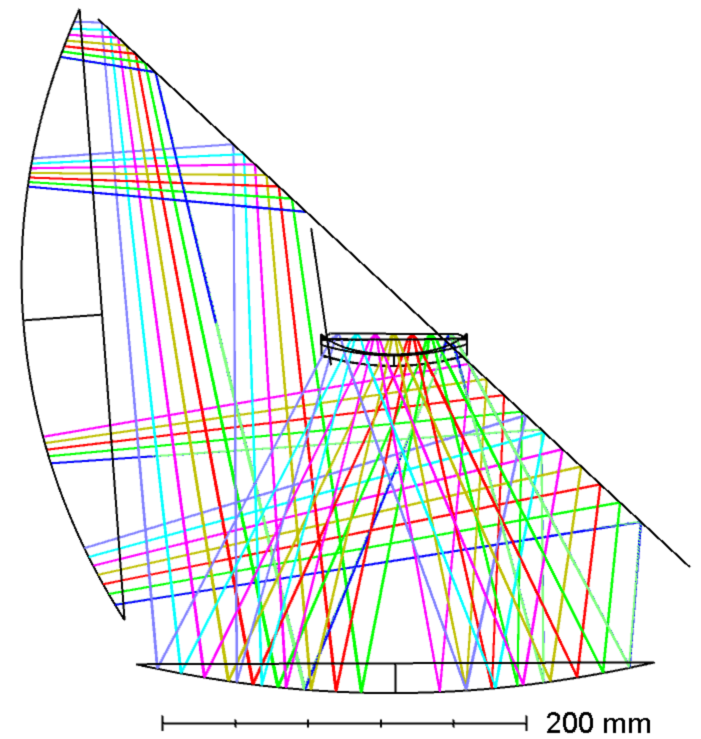}     &  & \includegraphics[width=0.35\linewidth]{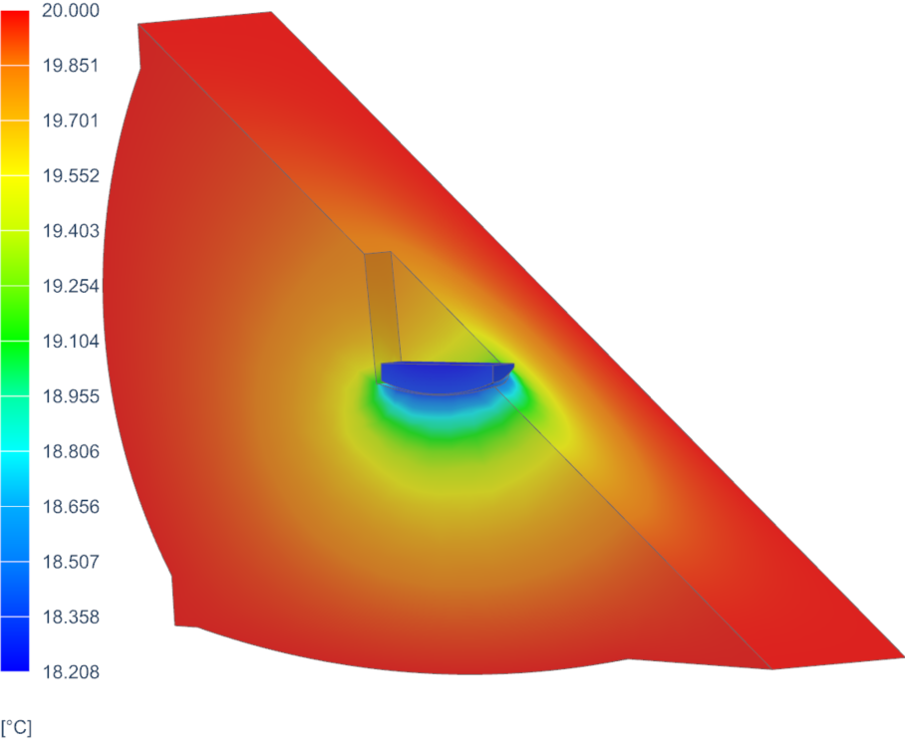}
    \end{tabular}
    \caption{Left: Optical design of the FSS camera (IZ-arm). Right: Thermal map of the FSS camera.}
    \label{fig:fss_cam}
\end{figure}

The other problem is a thermal issue in the camera. As the nose of the cryostat will be inside the block, this will generate a temperature gradient and therefore a refractive index gradient that could affect the image quality. A Finite Element Analysis (FEA) was conducted to obtain the temperature gradient at thermal equilibrium, the results of which are presented in Figure \ref{fig:fss_cam}.

But given the temperature differences in the glass, we end up with very small refractive index differences. The strongest contribution would be in the field lens. But given its proximity to the focal plane (1 $mm$), we assumed that there would be no impact on the image quality.

Beyond these thermal aspects, the FSS also innovates in the use of an uncommon material: Yttrium Aluminium Garnet (YAG). A crystalline solid often doped and used as an amplifying medium in lasers. But here, we have used it undoped for our field lens. Thanks to its refractive index ($n=1.833$), we can benefit from a gain in image quality that is not negligible.

\subsubsection{4x6}

Although innovative, the catadioptric camera of the FSS raises many questions, notably regarding feasibility, risk, and alignment. A dioptric camera would reduce all these risks. The 4x6 is a dioptric camera solution. Slower than the FSS with f/1.17, and with less impressive image quality with RMS $< d/6$. It has four arms named $UB$, $V$, $R$, and $IZ$, with 6 cm detectors.

With a spectral resolution $R > 3000$ everywhere and $R > 4000$ at 860nm, the spectral coverage and spectral resolution are given in detail in Table \ref{tab:4x6_cov}.

\begin{table}[h]
    \centering
    \begin{tabular}{l|c|c|c|c}
         & UB & V & R & IZ \\
    \hline
    $\lambda_{min}$ & 370 & 474 & 609 & 777 \\
    $\lambda_{max}$ & 490 & 627 & 805 & 980 \\
    $R_{min}$ & 3000 & 3000 & 3000 & 3614 \\
    $R_{max}$ & 3973 & 3968 & 3966 & 4558
    \end{tabular}
    \caption{Spectral coverage and spectral resolution in each arm for the 4x6.}
    \label{tab:4x6_cov}
\end{table}

\begin{figure}[H]
    \centering
    \begin{tabular}{cc}
    \includegraphics[width=0.4\linewidth]{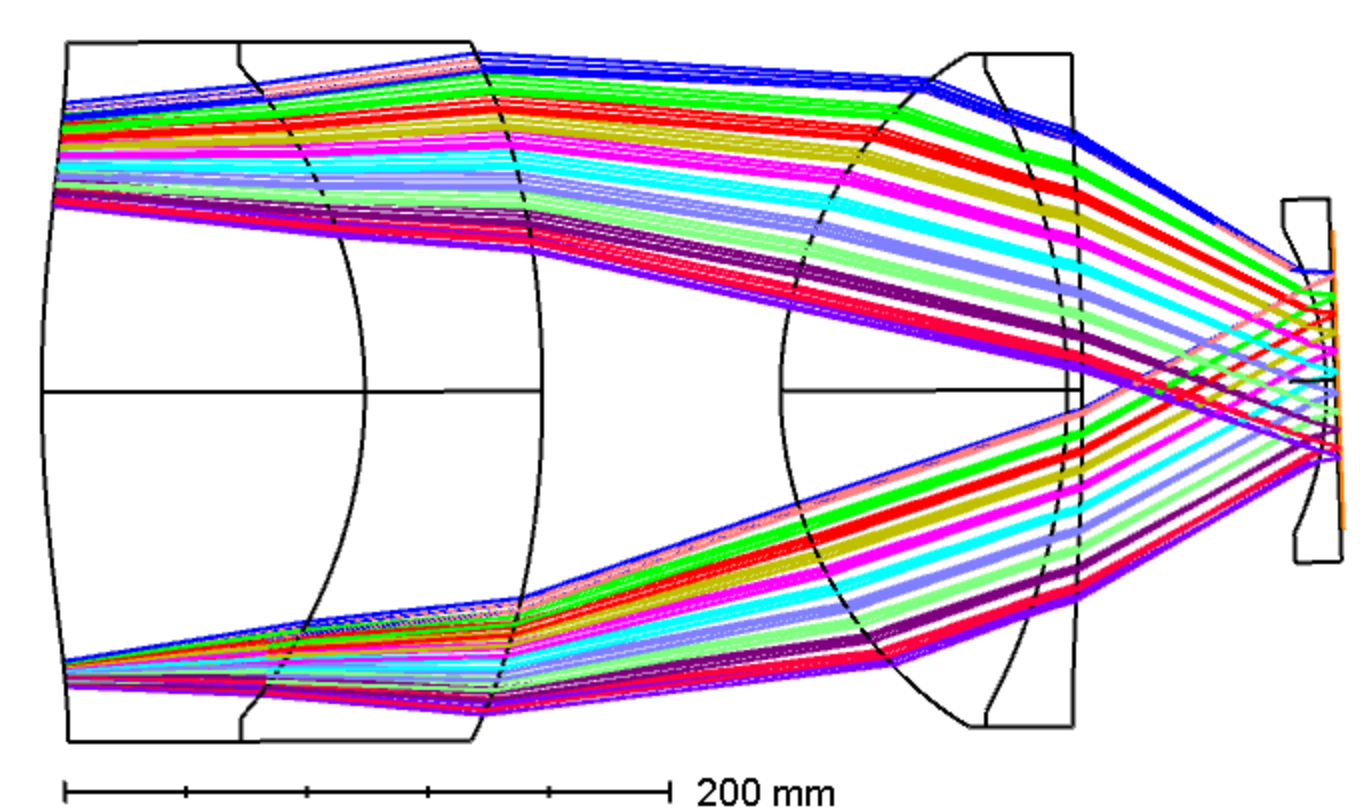}
    & \includegraphics[width=0.3916\linewidth]{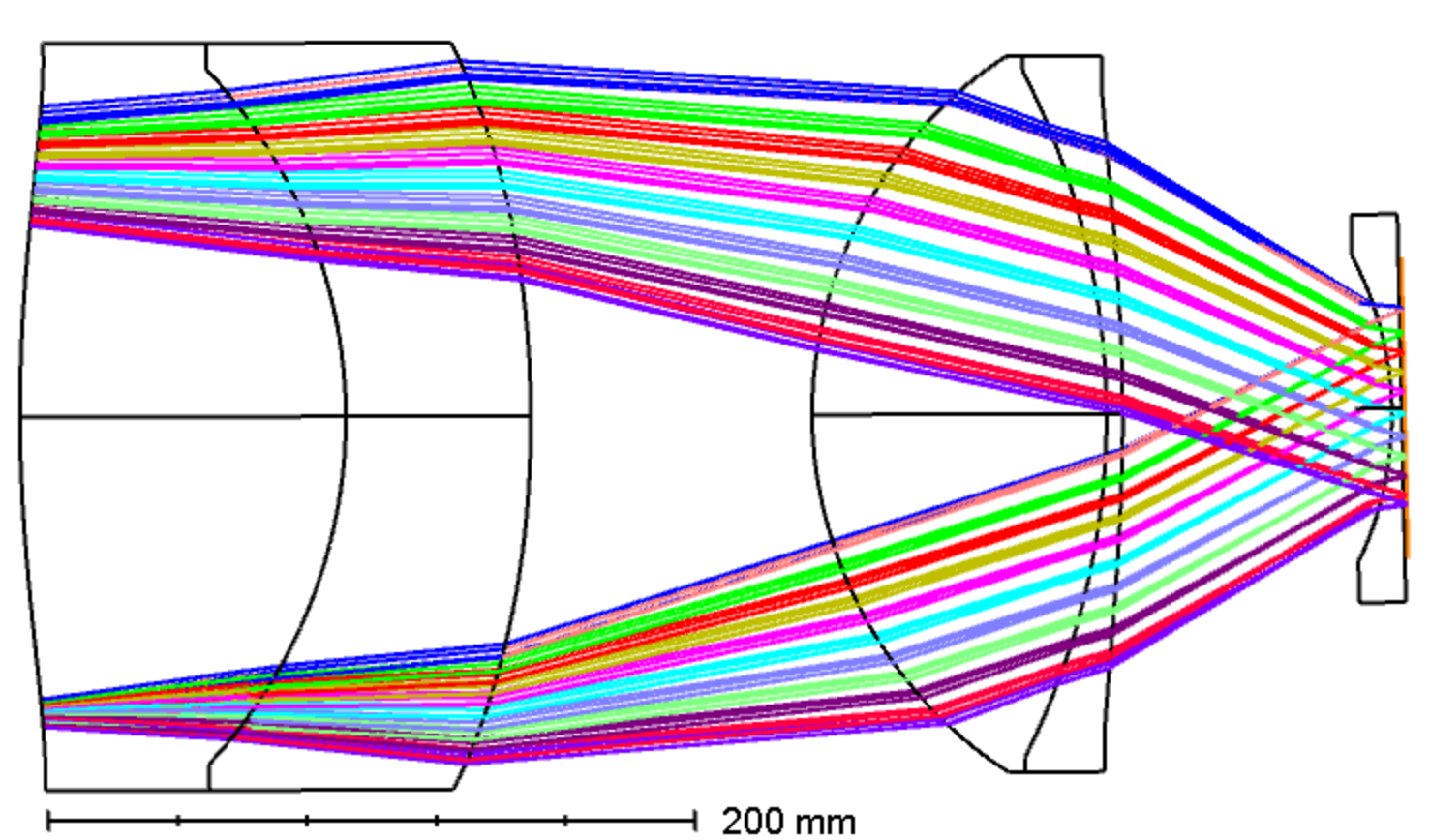}
    \end{tabular}
    \caption{4x6 camera of UB-arm (left) and IZ-arm (right).}
    \label{fig:4x6_cam}
\end{figure}

The cameras of the 4x6 are all composed of two achromatic doublets (Figure \ref{fig:4x6_cam}) but presenting strong asphericity. To overcome this problem, we use YAG field lenses which, as with the FSS, allow image quality to be improved and asphericity to be reduced.

\subsubsection{3x9}

\begin{figure}[h]
    \centering
    \begin{tabular}{cc}
    \includegraphics[width=0.4\linewidth]{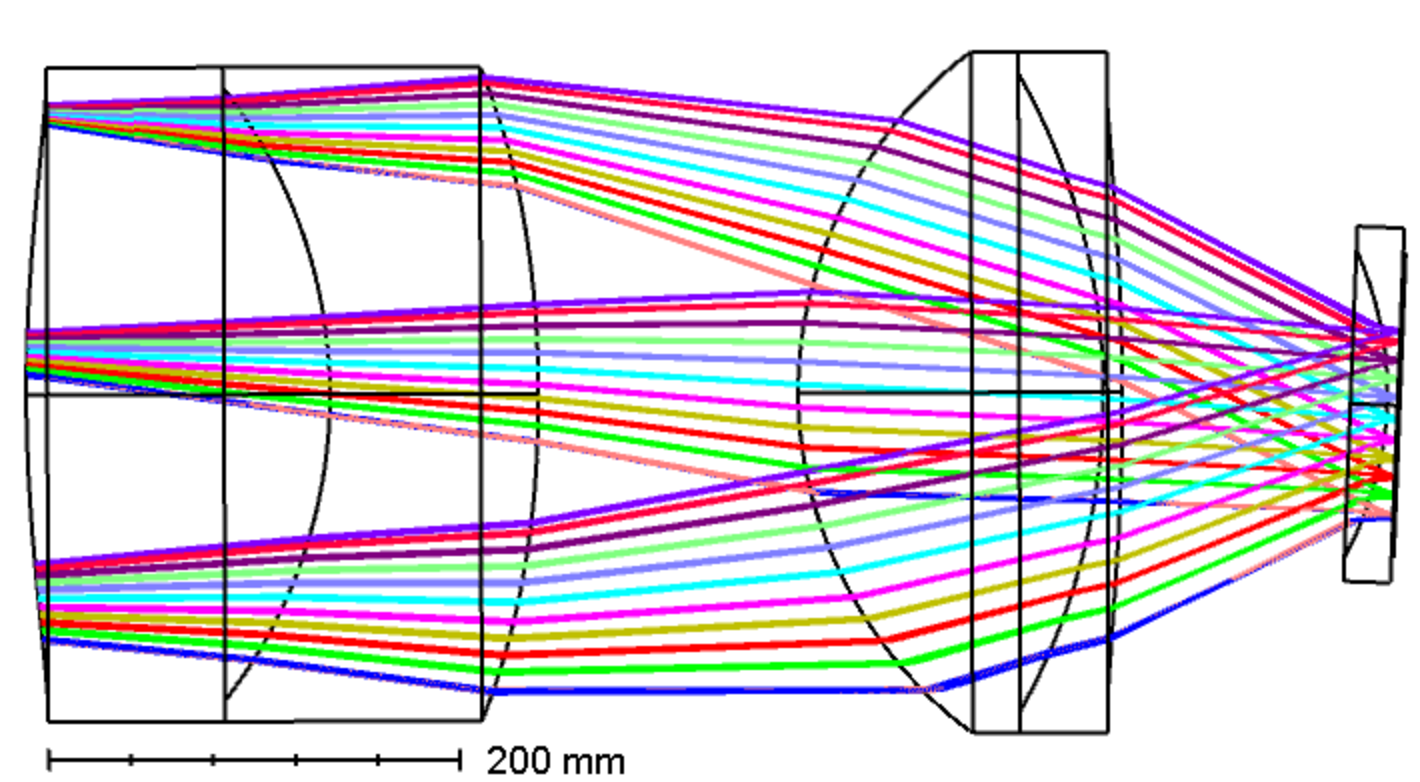}
    & \includegraphics[width=0.394\linewidth]{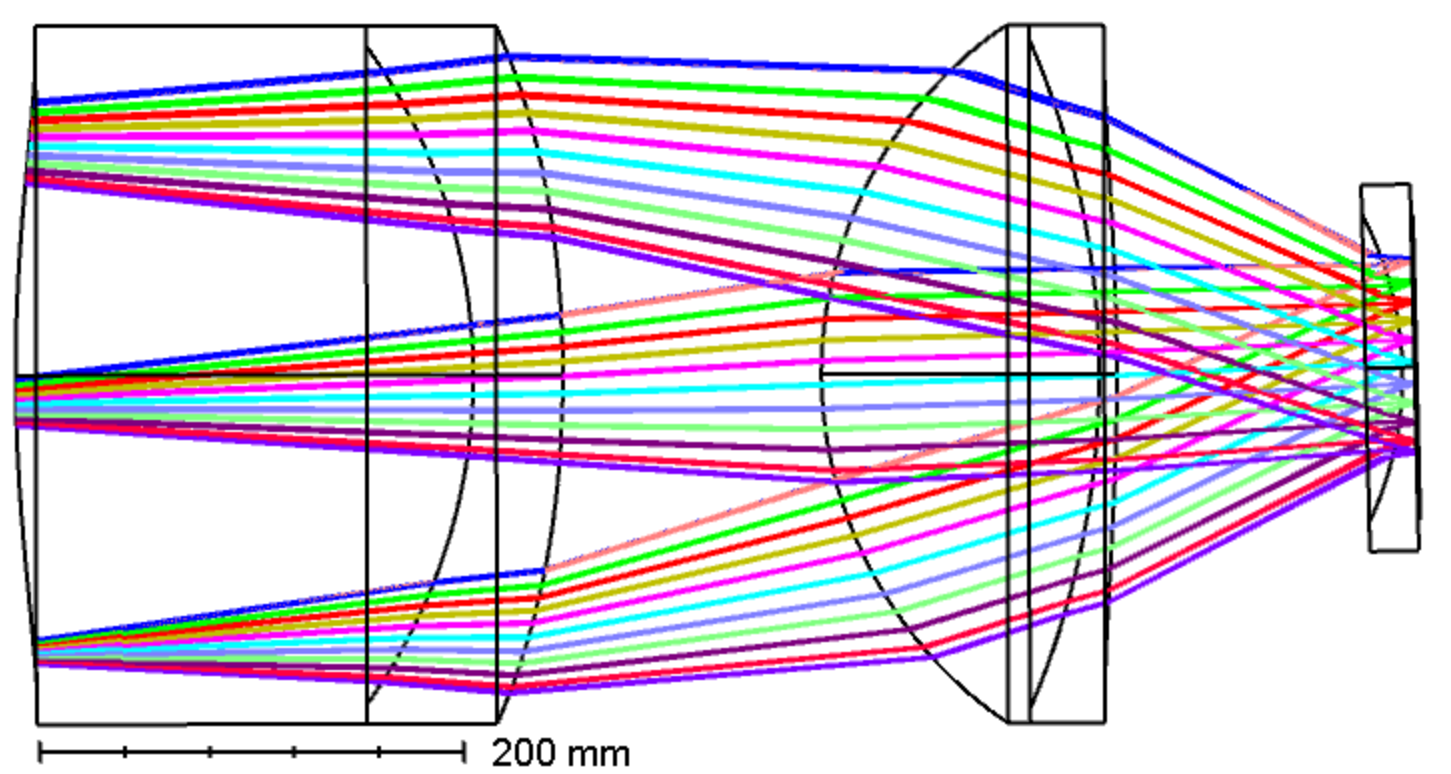}
    \end{tabular}
    \caption{3x9 camera of UV-arm (left) and IZ-arm (right).}
    \label{fig:3x9_cam}
\end{figure}

The 3x9 derives from the 4x6 by using the same type of camera, and by increasing the detector size to 9cm. The 3x9 allows for a faster camera at f/1.31, and achieves an RMS $<d/6$ while reducing the number of spectrographs. This three-arm design, named $UV$, $R$, and $IZ$, also has two achromatic doublets and YAG field lenses to reduce the asphericity of the achromatic doublets (Figure \ref{fig:3x9_cam}).

For the spectral resolution, we have $R > 3000$ everywhere and $R > 4000$ at 860 nm. The spectral coverage and spectral resolution are detailed in Table \ref{tab:3x9_cov}.

\begin{table}[h]
    \centering
    \begin{tabular}{l|c|c|c}
         & UV & R & IZ \\
    \hline
    $\lambda_{min}$ & 370 & 516 & 719 \\
    $\lambda_{max}$ & 532 & 740 & 980 \\
    $R_{min}$ & 3000 & 3000 & 3674 \\
    $R_{max}$ & 4313 & 4302 & 5014
    \end{tabular}
    \caption{Spectral coverage and spectral resolution in each arm for the 3x9.}
    \label{tab:3x9_cov}
\end{table}

A disadvantage of this design is the optical diameter. Going from $\sim$ 230 mm for the 4x6 to $\sim$ 330 mm. Which could considerably increase the production cost of this design.

\subsubsection{4x9}

\begin{figure}[H]
    \centering
    \begin{tabular}{cc}
    \includegraphics[width=0.4\linewidth]{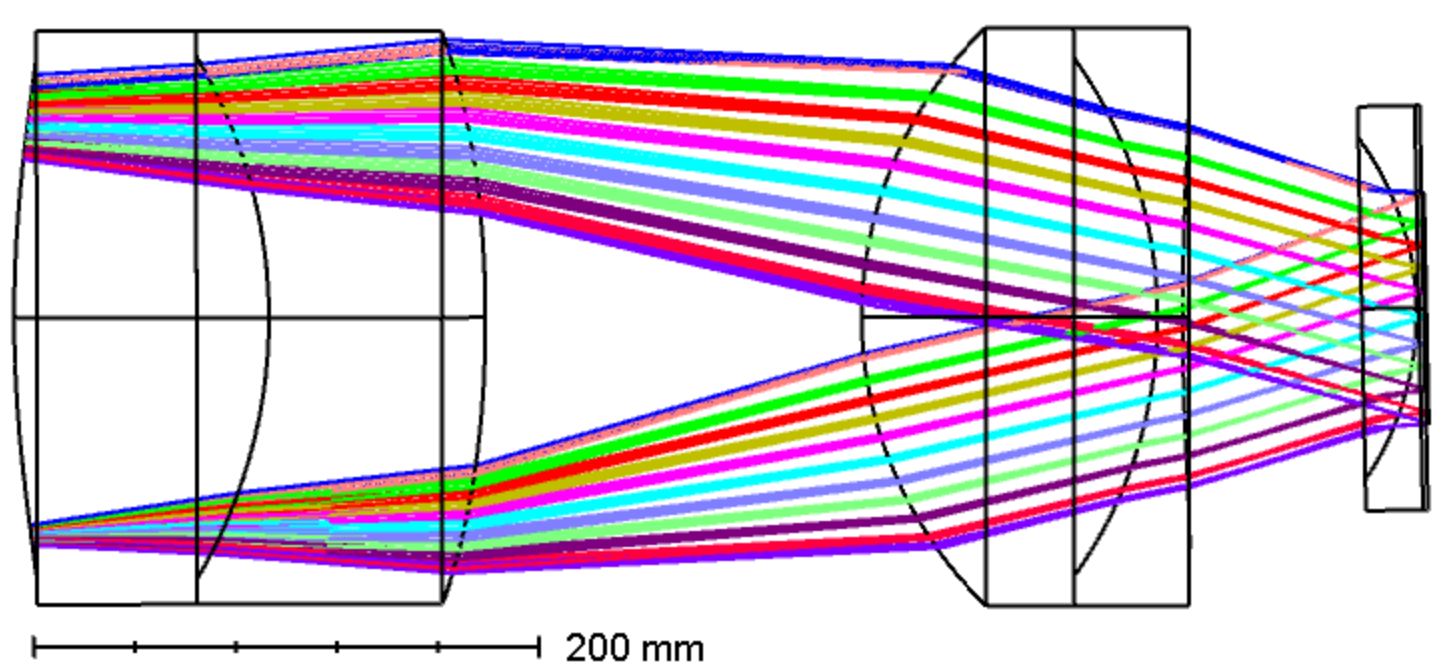}
    & \includegraphics[width=0.383\linewidth]{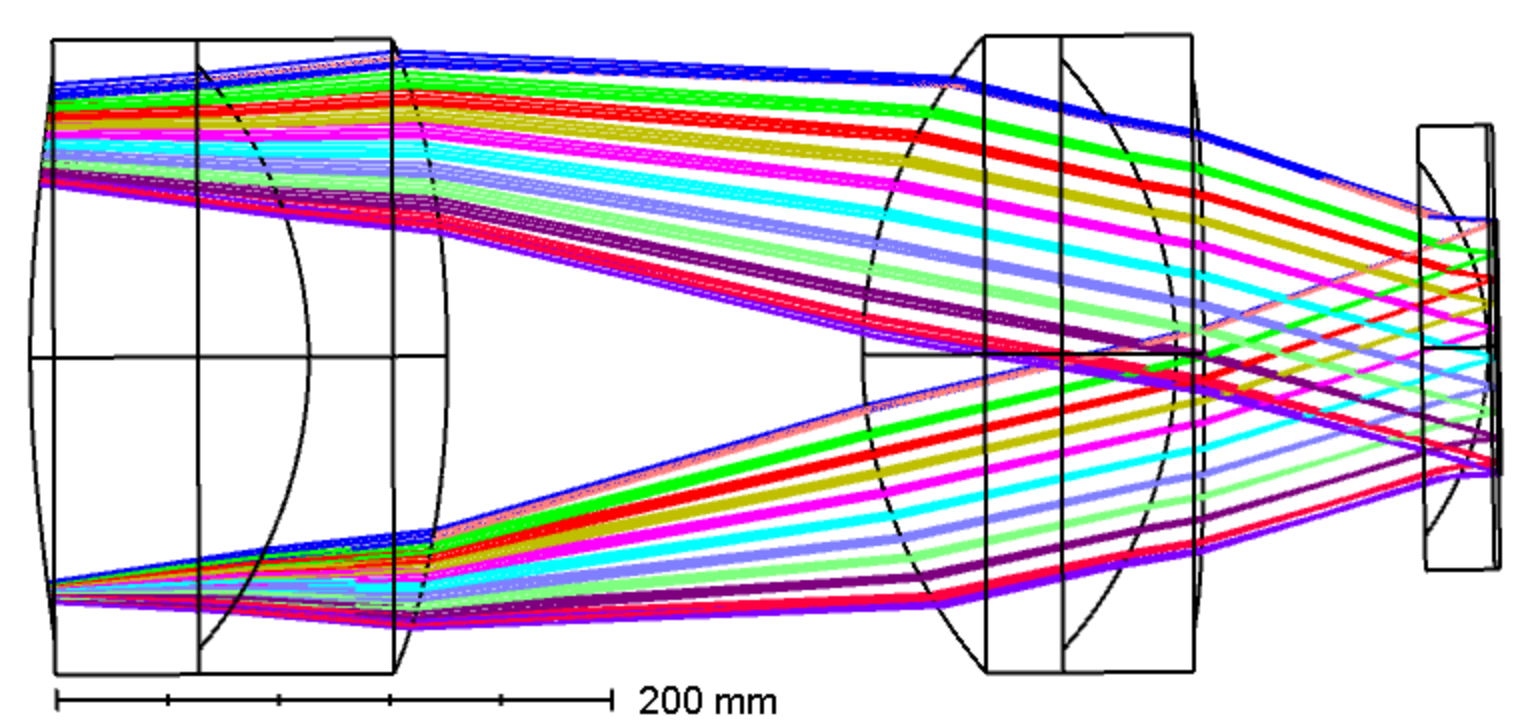}
    \end{tabular}
    \caption{4x9 camera of UB-arm (left) and IZ-arm (right).}
    \label{fig:4x9_cam}
\end{figure}

Another design derived from the 4x6 is the 4x9. By increasing the detector size to 9 cm, we reduce the asphericity of the optics (Figure \ref{fig:4x9_cam}). This allows us to revert to fused silica field lenses. The 4x9 then presents itself as the design with the least risk during production, with a slower camera at f/1.75 and image quality with RMS $<d/6$.

Like the 4x6, the 4x9 has four arms named $UB$, $V$, $R$, and $IZ$. It also has the same spectral resolution, namely $R>3000$ everywhere, and $R>4000$ at 860 nm. The spectral coverage and spectral resolution are detailed in Table \ref{tab:4x9_cov}. It should be noted that the 4x9 spectral coverage is not absolutely identical to that of the 4x6.

\begin{table}[h]
    \centering
    \begin{tabular}{l|c|c|c|c}
         & UB & V & R & IZ \\
    \hline
    $\lambda_{min}$ & 370 & 472 & 606 & 774 \\
    $\lambda_{max}$ & 489 & 624 & 802 & 980 \\
    $R_{min}$ & 3000 & 3000 & 3000 & 3600 \\
    $R_{max}$ & 3965 & 3966 & 3970 & 4558
    \end{tabular}
    \caption{Spectral coverage and spectral resolution in each arm for the 4x9.}
    \label{tab:4x9_cov}
\end{table}

Naturally, the 4x9 returns to optics with a diameter similar to the 4x6 ($\sim$ 230 mm). And less pronounced asphericity.

\subsubsection{A variant: 4x9Y}

\begin{figure}[h]
    \centering
    \begin{tabular}{cc}
    \includegraphics[width=0.4\linewidth]{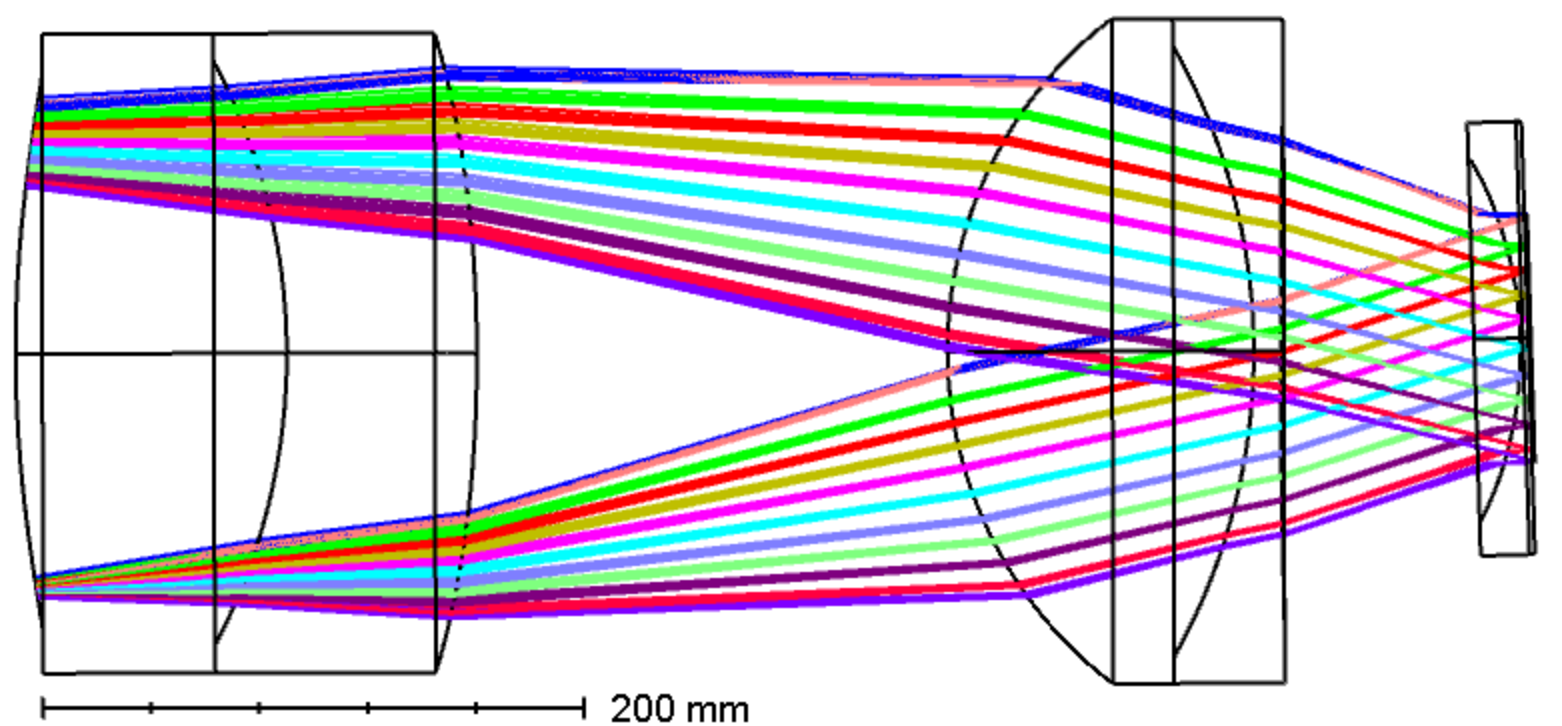}
    & \includegraphics[width=0.37\linewidth]{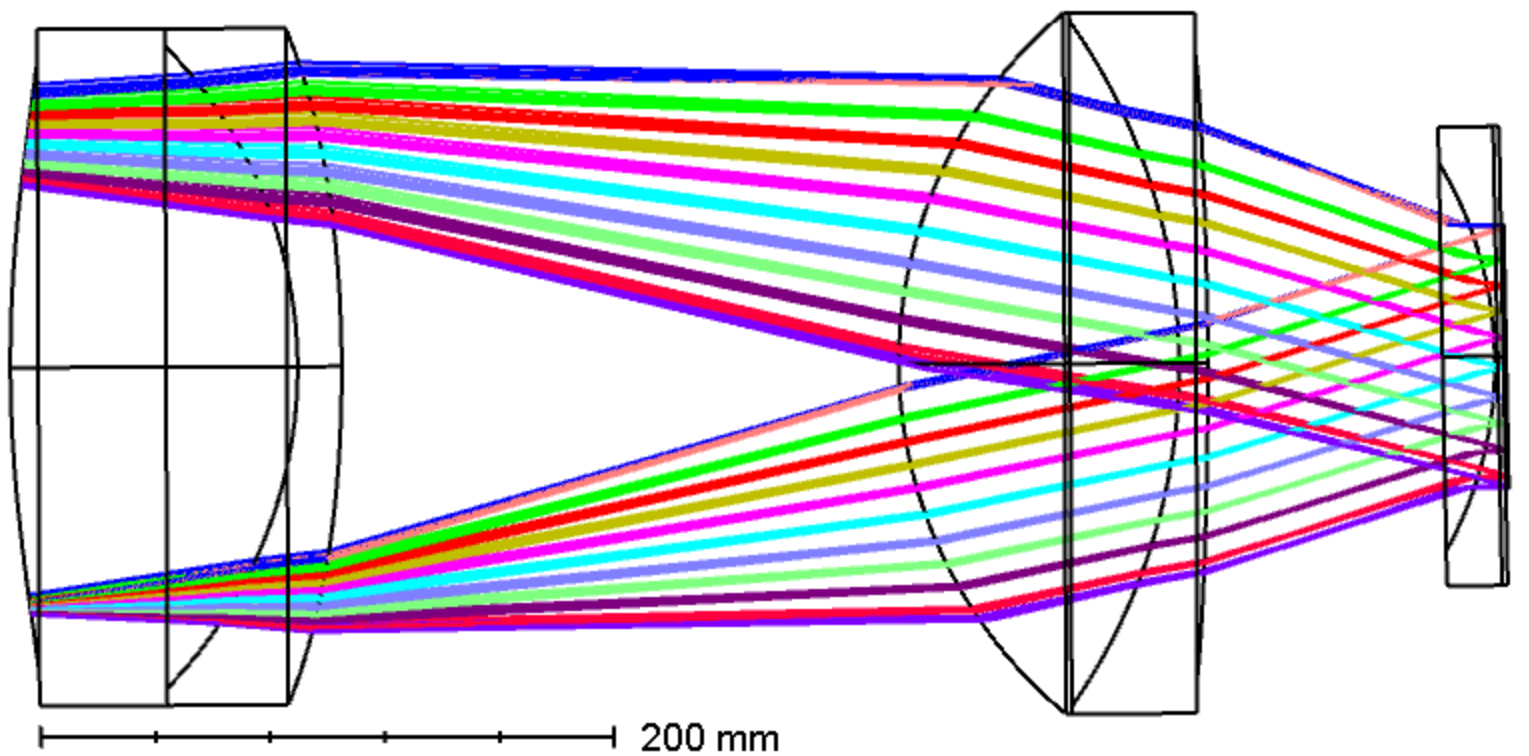}
    \end{tabular}
    \caption{4x9Y camera of UB-arm (left) and IZ-arm (right).}
    \label{fig:4x9Y_cam}
\end{figure}

It seems possible to us to further reduce the asphericity of the 4x9. To do this, we once again use YAG field lenses, making our cameras (Figure \ref{fig:4x9Y_cam}) faster in the process with f/1.60 and reducing the number of spectrographs to 53 units. The image quality reaches RMS $<d/6$. In comparison with the FSS (RMS $<d/10$), we obtain a similar image quality (see Section \ref{sec:tradeoff}).

We also have a gain in spectral resolution with $R>3200$ everywhere and $R>4300$ at 860 nm. The spectral coverage, meanwhile, remains unchanged (Table \ref{tab:4x9Y_cov}).

\begin{table}[h]
    \centering
    \begin{tabular}{l|c|c|c|c}
         & UB & V & R & IZ \\
    \hline
    $\lambda_{min}$ & 370 & 472 & 606 & 774 \\
    $\lambda_{max}$ & 489 & 624 & 802 & 980 \\
    $R_{min}$ & 3200 & 3200 & 3200 & 3870 \\
    $R_{max}$ & 4229 & 4231 & 4235 & 4900
    \end{tabular}
    \caption{Spectral coverage and spectral resolution in each arm for the 4x9Y.}
    \label{tab:4x9Y_cov}
\end{table}

However, this last design slightly increases the lens diameter, the largest of which are $\sim$ 250 mm. This may again have an impact on the production cost, but the effect will be felt much less than with the 3x9.

\subsection{YAG properties}

\begin{figure}[H]
    \centering
    \includegraphics[width=0.8\linewidth]{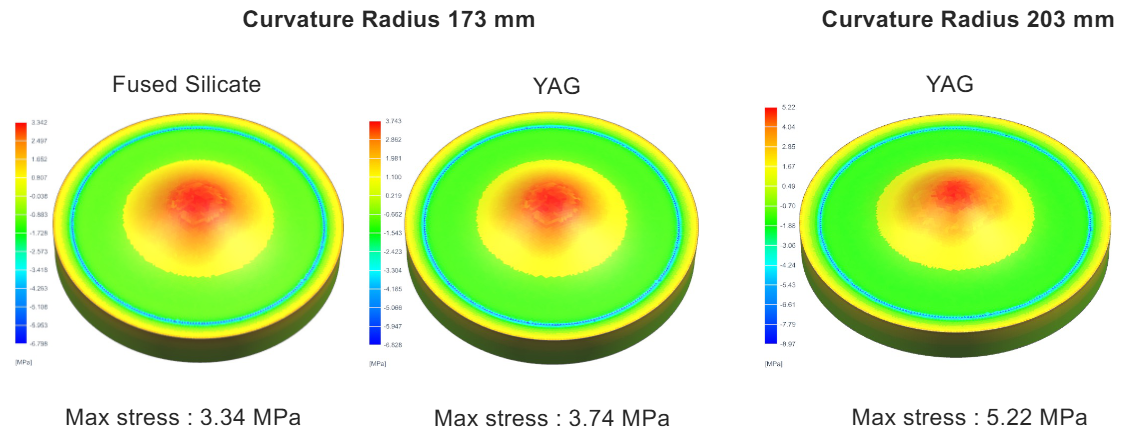}
    \caption{Mechanical stress on different field lens. Biggest values (min) are singularities from the boundary conditions.}
    \label{fig:stress}
\end{figure}

With the 4x9 and 4x9Y, we can run FEA simulations to test some properties. Both designs have field lenses with different radii of curvature in each arm, making a direct comparison difficult. Fortunately, both designs share a field lens with the same radius of curvature. Figure \ref{fig:stress} shows the results for these two field lenses, as well as the mechanical stress that the YAG field lens with the largest radius of curvature can withstand.

These results show that a YAG field lens is more robust. This mechanical stress will be applied to the field lenses since the cryostats will be directly fixed onto them in order to place the detectors under vacuum and at temperature.

\subsection{Detector properties}

WST is planned for the 2040s. Today, CCDs face an uncertain future: they are increasingly difficult to source and come at a high cost. It therefore makes sense to use CMOS detectors as our reference. They are less expensive, easier to obtain, and have high technological potential. CMOS detectors could achieve low readout noise and would allow us to bin pixels at readout.

\begin{figure}[h]
    \centering
    \begin{tabular}{cc}
        \includegraphics[width=0.54\linewidth]{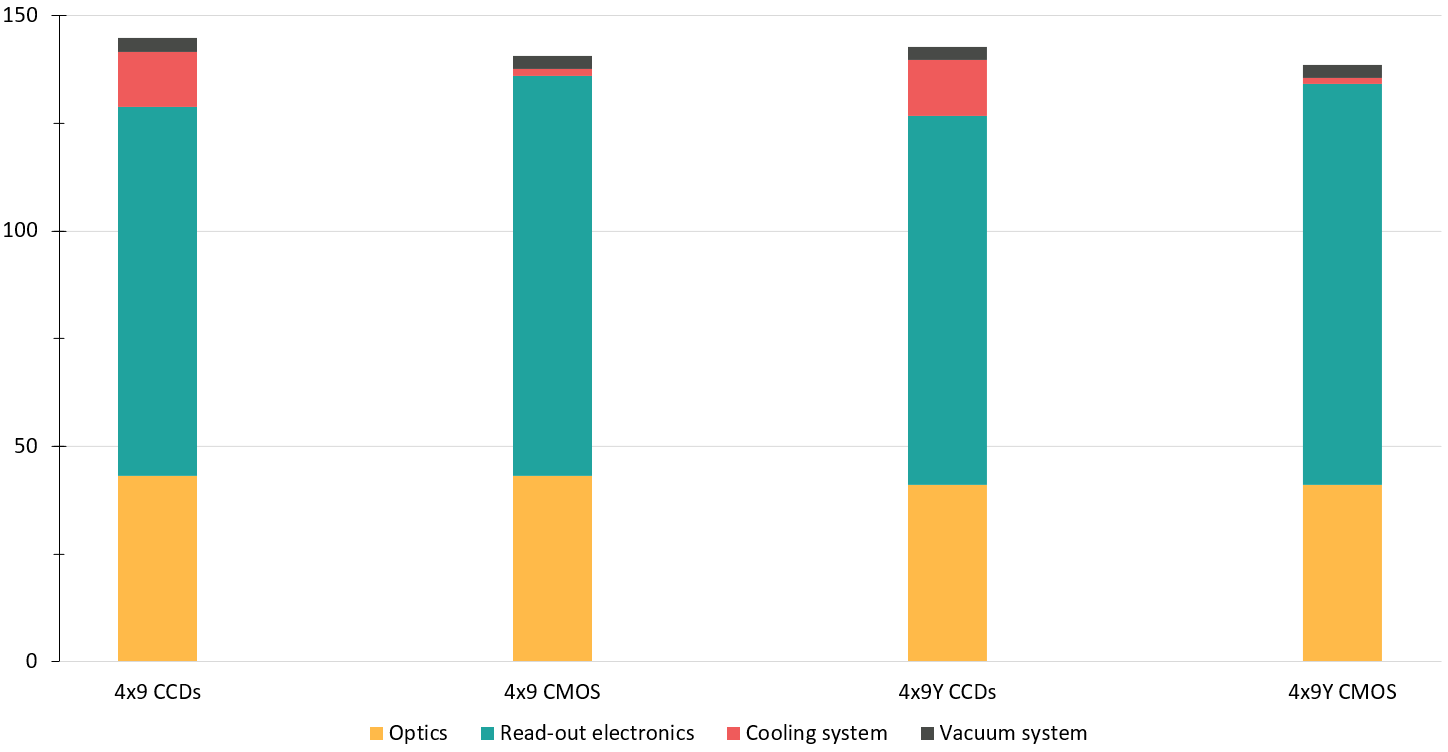} &
        \includegraphics[width=0.40\linewidth]{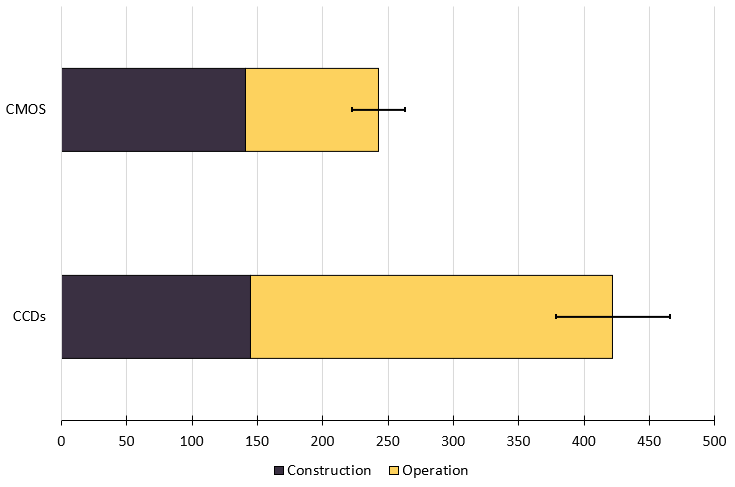}
    \end{tabular}
    \caption{Left: Distribution per components of the construction footprint for the four different MOS-LR designs. Right: Construction and 1-year operation carbon footprint for the 4x9 option with CCDs and CMOS (in tCO2eq).}
    \label{fig:carbon_footprint}
\end{figure}

Another point of view is the impact of the carbon footprint during the trade-off. In parallel of the trade-of study, a study was conducted [\citenum{Freour2026}]. This shows that the carbon footprint generated by the construction of the two designs is almost identical (Figure \ref{fig:carbon_footprint}). In fact, the difference is mainly felt through the choice of CCD or CMOS detector, whose electricity and cooling requirements differ.

Beyond the CCD vs CMOS aspect, we also want to study the impact of pixel size (10 or 15 $\mu m$) on the survey. For this, we look at the peak count level and the survey speed hit in the UB arm.

The peak count level indicates the maximum signal recorded per pixel, expressed in ADU. Smaller pixels (10 µm) collect proportionally less flux than larger ones (15 µm), providing a wider margin against saturation but a lower signal-to-noise ratio per pixel.

The survey speed hit quantifies the reduction in observing efficiency induced by the detector's read noise relative to the signal collected. Smaller pixels, with a reduced photon collection area, require longer integration times to reach the target SNR, resulting in a larger penalty on overall survey speed.

\begin{table}[h]
    \centering
    \begin{tabular}{l|c|c|c|c|c|c|c|c|c|c}
            & \multicolumn{2}{|c|}{FSS} & \multicolumn{2}{|c|}{3x9} & \multicolumn{2}{|c|}{4x6} & \multicolumn{2}{|c|}{4x9}  & \multicolumn{2}{|c}{4x9Y}\\
            & 10 $\mu m$ & 15 $\mu m$ & 10 $\mu m$ & 15 $\mu m$ & 10 $\mu m$ & 15 $\mu m$ & 10 $\mu m$ & 15 $\mu m$ & 10 $\mu m$ & 15 $\mu m$ \\
        \hline
        Peak (counts) & 21 & 48 & 8 & 17 & 10 & 22 & 4 & 10 & 4 & 10 \\
        Hit (\%) & 3 & 2 & 8 & 4 & 6 & 3 & 10 & 6 & 10 & 6 \\ 
        Binning (px) & 3.78 & 2.52 & 6.19 & 4.12 & 5.53 & 3.68 & 8.27 & 5.51 & 7.56 & 5.04
    \end{tabular}
    \caption{Peak and hit survey for 10 and 15 $\mu m$ pixel size.}
    \label{tab:peak_and_hit}
\end{table}

Thanks to these results (Table \ref{tab:peak_and_hit}), we select the 15 $\mu m$ pixel size as the reference. Although it is consistent with other MOS facilities (Table \ref{tab:mos_multiplex}), we were wondering whether the use of a CMOS could change this trend.

\begin{equation}
    \text{Binning} = d_{\text{fib}} \frac{\sqrt{3}}{2} \frac{F_{\text{cam}}}{F_{\text{coll}}} \frac{1}{dx_{\text{pix}}}\cdot \frac{\sqrt{\pi}}{2}
    \label{eq:bin}
\end{equation}

Apart from the low readout noise of CMOS detectors, these can also allow us to bin pixels at readout. To evaluate this binning factor, we first calculate the pixel sampling denoted $k_{\text{pix}}$ (Equation (\ref{eq:k_pix})). The binning factor is obtained with Equation (\ref{eq:bin}).

Finally, we require a distance between the field lens and the focal plane of approximately 4$mm$. The trade-off will be carried out at a distance of 1$mm$ and the selected design will be re-optimised for a distance of 3.5$mm$. Section \ref{sec:evol} will detail the consequence of this simultaneous change alongside the change of requirements (Section \ref{sec:requir}).

\section{TRADE-OFF}
\label{sec:tradeoff}

In total, we have therefore produced five optical designs meeting all the requirements.

For our trade-off, we will compare the performance of each of these designs. Namely the image quality, the throughput, and the spectral resolution.

Each of these designs has a different footprint area, volume, and number of spectrographs. These being a constraint, from a logistical point of view, on their installation.

Beyond this, we must also build a model allowing us to evaluate their cost and their risk with regard to the production of the spectrographs.

Finally, we wish to compare the environmental impact of their construction and during operation.

\subsection{Image quality}

\begin{figure}[H]
    \centering
    \includegraphics[width=0.5\linewidth]{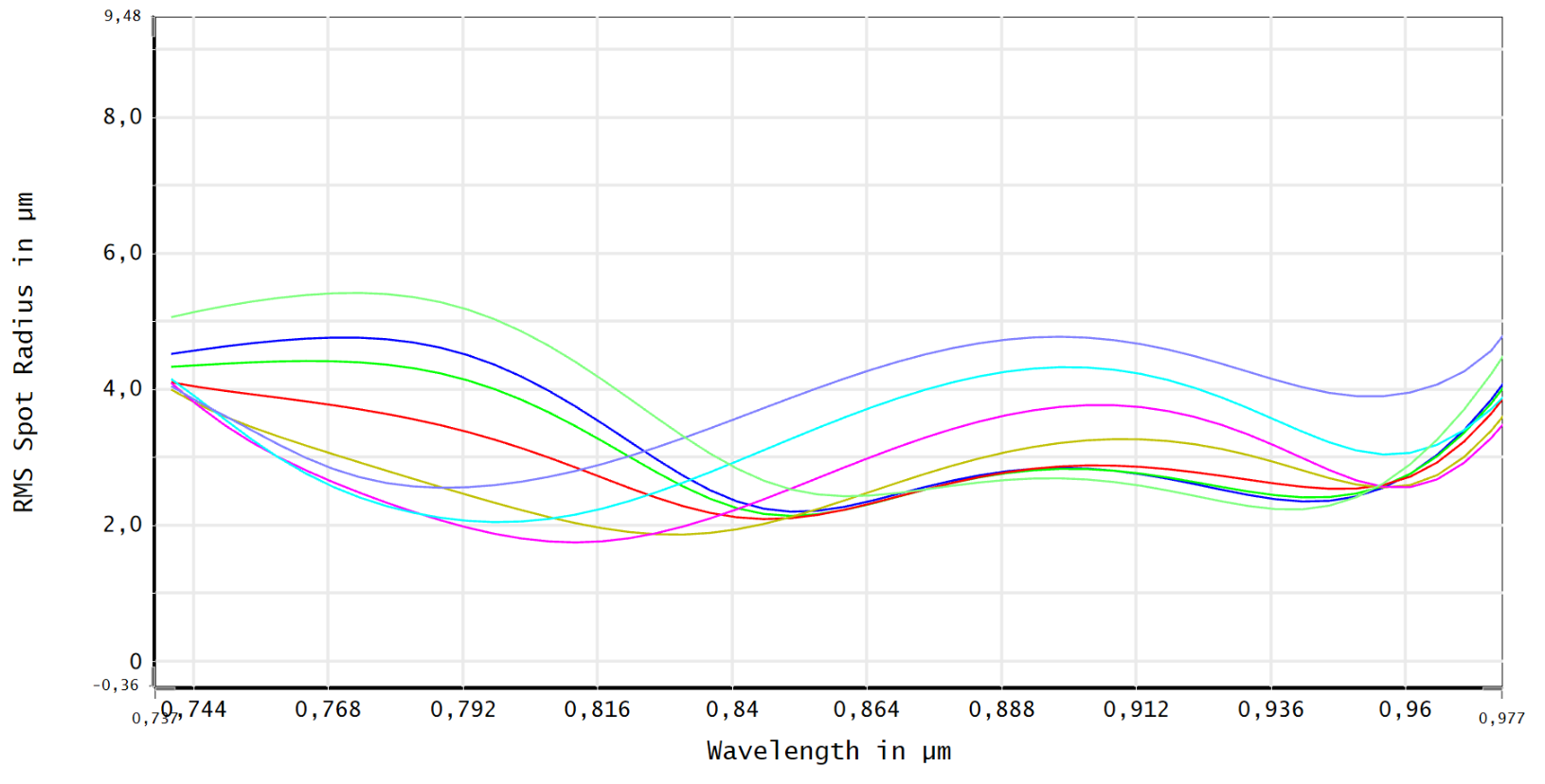}
    \caption{FSS IZ-arm RMS spot radius image quality}
    \label{fig:fss_rms_iz}
\end{figure}

To study the image quality, we use the RMS spot radius as a function of wavelength. As in Figure \ref{fig:fss_rms_iz} which is an example for the FSS in the IZ arm. So we have as many graphs as there are arms for each design.

\begin{figure}[t]
    \centering
    \includegraphics[width=1.0\linewidth]{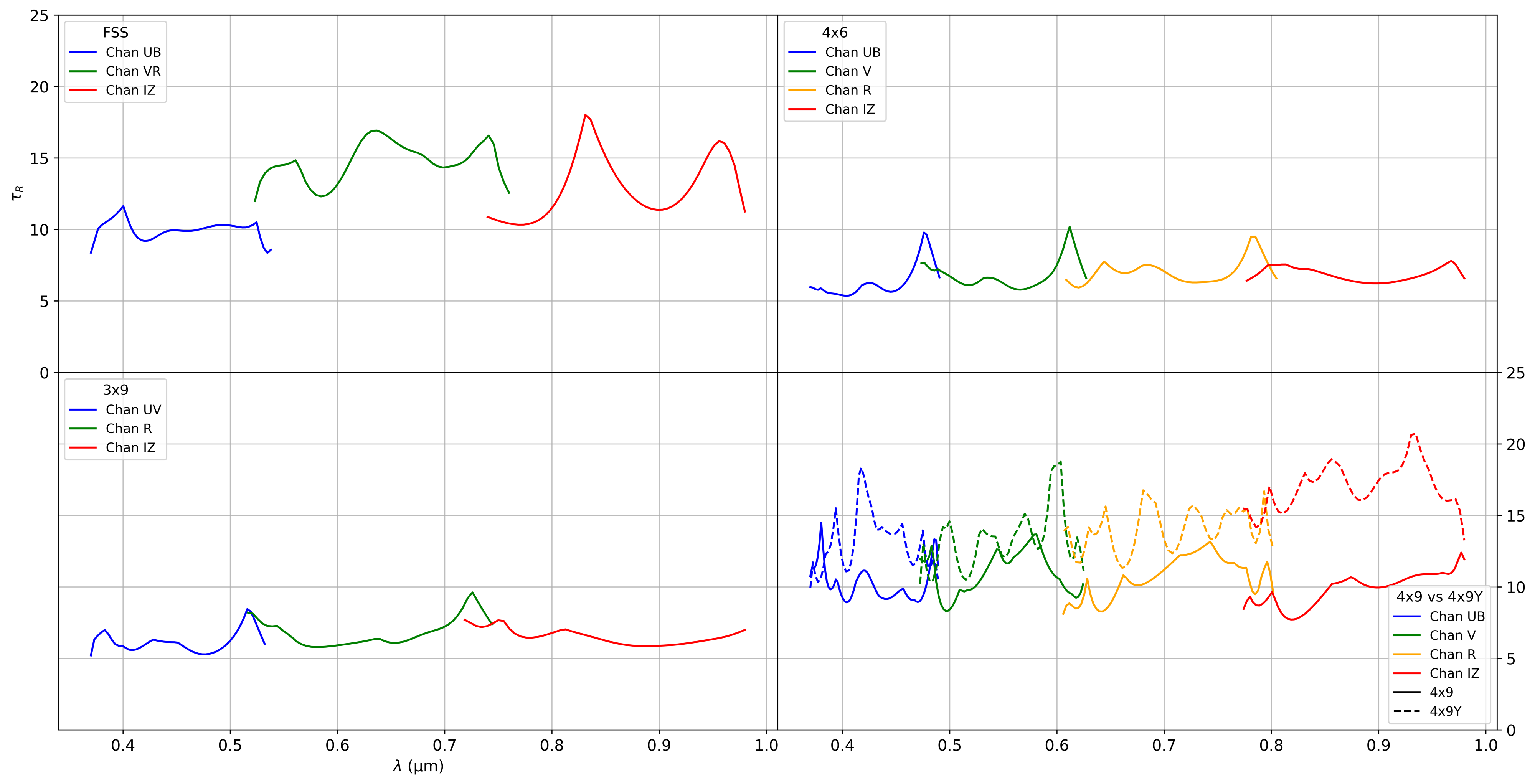}
    \caption{Image quality: ratio between the projected fibre diameter and the RMS over 90\% FoV.}
    \label{fig:RMS}
\end{figure}

Given the number of designs, this gives us eighteen graphs to study. To simplify our analysis, we use 90\% of the field of view (FoV) (i.e. 90\% of the slit length) to obtain the RMS. We then take the maximum RMS value over all fields to obtain a single curve per arm. We then express the ratio $\tau_R$ between the projected fibre diameter and the RMS. And finally, we group the curves onto a single graph for each design (Figure \ref{fig:RMS}).

Regarding the 4x9 and the 4x9Y, it seemed interesting to us to plot their curves on the same graph in order to better understand the performance gain of the 4x9Y.

The worst and best values of these graphics, as well as the projected fibre size, are detailed in Table \ref{tab:IQ}. These two values obtained for the RMS will be used in the calculation of the scores.

\begin{table}[h]
    \centering
    \begin{tabular}{l|c|c|c|c|c}
         & FSS & 3x9 & 4x6 & 4x9 & 4x9Y \\
         \hline
         Projected fibre diameter ($\mu m$) & 49.21 & 80.62 & 72.00 & 107.69 & 98.46 \\
         Best $\tau_\text{R}$ & 18.02 & 9.62 & 10.20 & 14.50 & 20.72 \\
         Worst $\tau_\text{R}$ & 8.37 & 5.21 & 5.36 & 7.72 & 9.92
         
    \end{tabular}
    \caption{Image quality for 90\% field of view (FoV).}
    \label{tab:IQ}
\end{table}

\subsection{Throughput}

For the throughput, we based ourselves on the anti-reflection data from the Laboratoire des Matériaux Avancés (LMA) for the lens or the collimator transmission factors.

\begin{equation}
    T (\lambda) = \frac{T_0}{1 + e^{\pm \frac{\lambda - \lambda_c}{\lambda_c \cdot \frac{\delta \lambda}{2}}}}
    \label{eq:dich}
\end{equation}

Regarding the dichroics, we use a model based on a sigmoid function defined in Equation (\ref{eq:dich}). Where $\delta\lambda$ is the dichroic overlap and $\lambda_c$ is the cut-off wavelength. That is to say, at this wavelength we will have a throughput of 50\%.

\begin{figure}[h]
    \centering
    \begin{tabular}{cc}
          \includegraphics[width=0.4\linewidth]{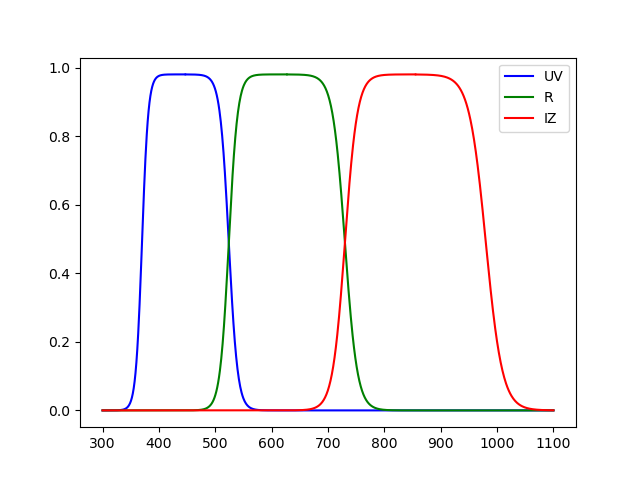}
        & \includegraphics[width=0.4\linewidth]{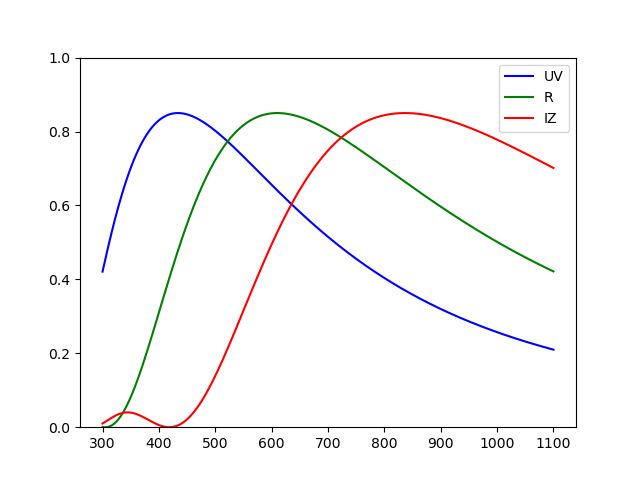}
    \end{tabular}
    \caption{Dichroïcs (left, $T_0=0.98$) and grating (right, $T_0 = 0.85$) throughput for 3x9 design.}
    \label{fig:3x9_dichs_and_grating}
\end{figure}

When we are at the lower bound of a band, we use the '$+$' sign. And when we are at the upper bound we use the '$-$' sign. The sign change is performed at the middle of the band. As an example, we obtain Figure \ref{fig:3x9_dichs_and_grating} (left) for the dichroic transmission of the 3x9.

\begin{equation}
    T(\lambda) = T_0 \cdot sinc^2(\pi (\sigma \lambda_B - 1))
    \label{eq:grating}
\end{equation}

Regarding the grating, we use Equation (\ref{eq:grating}). Where $\sigma = \frac{1}{\lambda}$ is the wavenumber. And $\lambda_B = \frac{2}{\sigma_{\inf} + \sigma_{\sup}}$ is the Blaze wavelength. Which then allows us to obtain Figure \ref{fig:3x9_dichs_and_grating} (right).

\begin{figure}[h]
    \centering
        \includegraphics[width=0.9\linewidth]{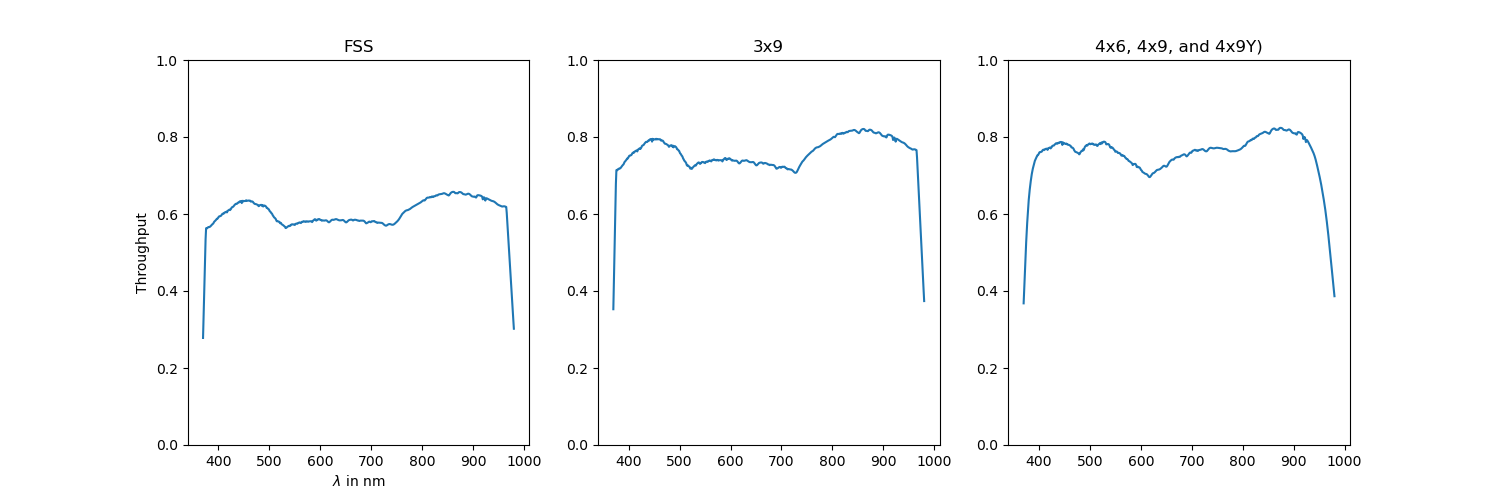}
    \caption{Throughput estimated.}
    \label{fig:Throughput}
\end{figure}

We then retrieve the average throughput for each optical concept, extract from the total throughput (Figure \ref{fig:Throughput}), which is detailed in Table \ref{tab:Average_throughput}.

\begin{table}[H]
    \centering
    \begin{tabular}{l|c|c|c|c|c}
         & FSS & 3x9 & 4x6 & 4x9 & 4x9Y \\
    \hline
    Average throughput (\%) & 59.5 & 74.4 & 75.2 & 75.2 & 75.2 
    \end{tabular}
    \caption{Average throughput.}
    \label{tab:Average_throughput}
\end{table}

\subsection{Spectral properties}

\begin{figure}[H]
    \centering
    \includegraphics[width=1.0\linewidth]{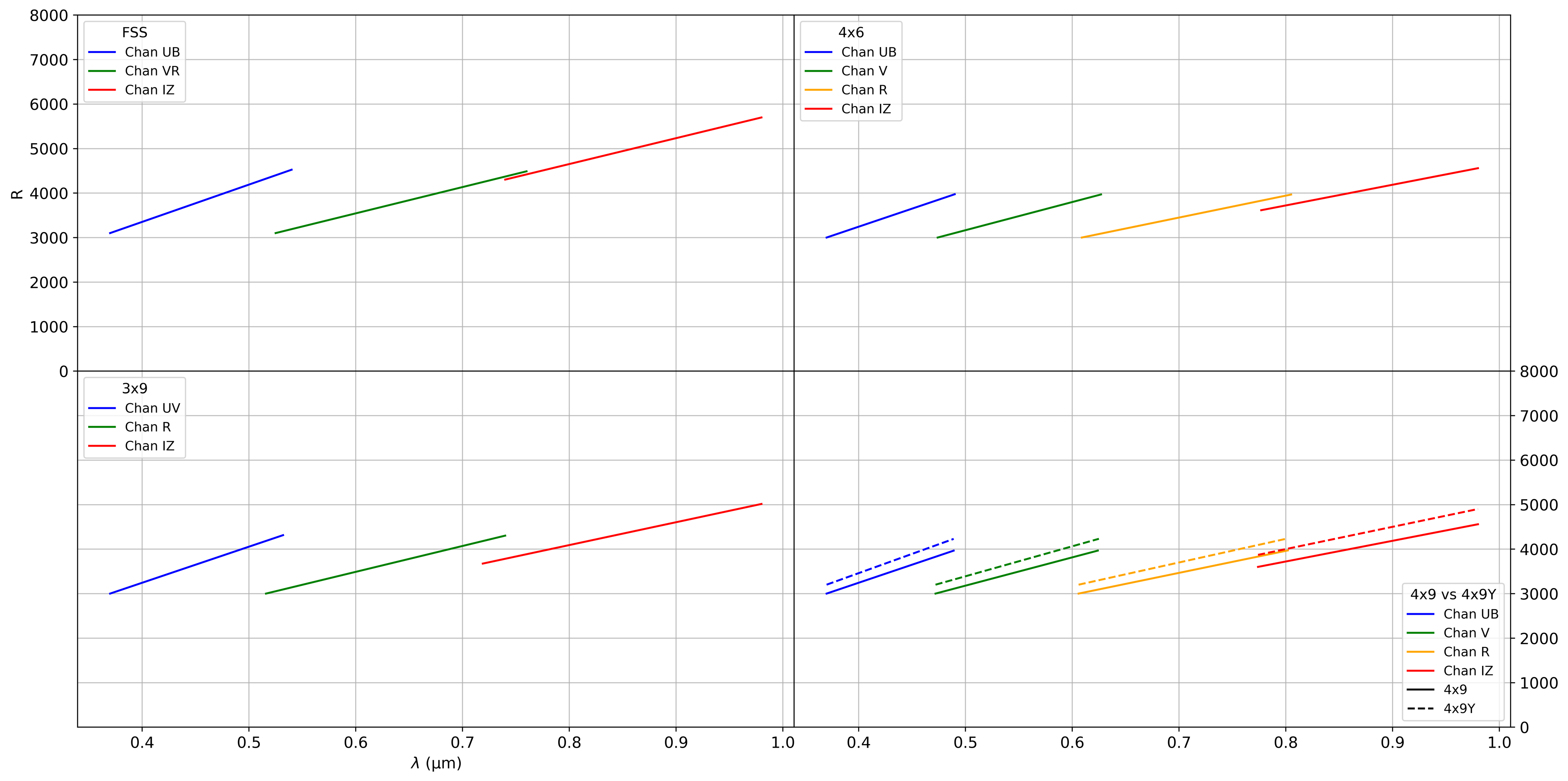}
    \caption{Spectral resolution for each design.}
    \label{fig:R}
\end{figure}

To compare the spectral resolution, we can look at the worst and best overall resolution (i.e. regardless of the arm). It is more interesting to look at the spectral resolution at the limits, at 370 and 980 nm, and at the lower limit of the IZ-arm in each design. The spectral resolution curves are given in Figure \ref{fig:R}. And the values of interest we have just mentioned are detailed in Table \ref{tab:R_LR}.

\begin{table}[h]
    \centering
    \begin{tabular}{l|c|c|c|c|c}
         & FSS & 3x9 & 4x6 & 4x9 & 4x9Y \\
    \hline
    $R_{min}$ at 370 $nm$ & 3100 & 3100 & 3000 & 3000 & 3200 \\
    $R_{min}$ in channel IZ & 4302 & 3837 & 3614 & 3600 & 3870 \\
    $R_{max}$ at 980 $nm$ & 5698 & 5014 & 4558 & 4558 & 4900
    \end{tabular}
    \caption{Spectral resolution at the limits and in channel IZ.}
    \label{tab:R_LR}
\end{table}

\subsection{Volume \& footprint}

To size the spectrographs, we take the maximum distance along the three axes (x, y, z). These distances contain only the optics of the designs. The cooling systems and the electronic systems are not included.

It also seemed interesting to us to compare the footprint area for bench integration (which is done conventionally). And for rack integration, a concept which will be presented in Section \ref{sec:evol}. The values of interest are detailed in Table \ref{tab:V_S}.

For the scoring, we used the volume and the footprint for a rack. Which seemed to us to be the best option. The constraints related to this concept will be discussed in Section \ref{sec:constraints}.

\begin{table}[h]
    \centering
    \begin{tabular}{l|c|c|c|c|c}
        & FSS & 3x9 & 4x6 & 4x9 & 4x9Y \\
    \hline
    Volume in $m^3$ & 2.57 & 2.55 & 0.87 & 0.98 & 0.95 \\
    Bench footprint in $m^2$ & 2.92 & 3.19 & 1.50 & 1.75 & 1.70 \\
    Rack footprint in $m^2$ & 1.01 & 1.04 & 0.56 & 0.58 & 0.58    
    \end{tabular}
    \caption{Volume and footprint for one spectrograph.}
    \label{tab:V_S}
\end{table}

\subsection{Cost \& Risk}

Evaluating the cost of a spectrograph is a tedious task. Establishing a cost model without certainty about its consistency is not necessarily the best way to constrain it.

\begin{equation}
    Cost \propto N_{\text{spec}} \sum_{i,j} D_{i,j}^2
    \label{eq:cost}
\end{equation}

A first-order approach would be to sum the squared diameters of all lenses (j) contained in each arm (i), and multiply the result by the number of spectrographs, ($N_{\text{spec}}$) like in Equation (\ref{eq:cost}). This provides a single manufacturing driver that accounts for both the number of optical elements to be manufactured and their total effective optical area, which is particularly relevant when ordering optical blanks and estimating polishing requirements.

Also, the risk was evaluated arbitrarily internally, taking into account the lens diameters, their shapes, feasibility, and material availability. The details are presented in Table \ref{tab:c&r}.

\begin{table}[h]
    \centering
    \begin{tabular}{l|c|c|c|c|c}
         & FSS & 3x9 & 4x6 & 4x9 & 4x9Y \\
    \hline
    Cost driver ($\propto m^2$) & 159.89 & 110.12 & 89.48 & 88.54 & 83.75 \\
    Risk & 5 & 4 & 2.5 & 2 & 2.5
    \end{tabular}
    \caption{Summary of the Cost \& Risk drivers.}
    \label{tab:c&r}
\end{table}

\subsection{Environmental Impact}

To evaluate the environmental impact, two periods must be distinguished. The construction period and the operation period. The construction period corresponds to the period of manufacturing and assembly of the optics to obtain a spectrograph. The operation period corresponds to the period during which the spectrographs are on site, already installed, tested, and commissioned. In other words, we are in the survey phase.

Estimating the carbon footprint of such a project is complex. In the same way as for the Cost \& Risk, we will compare elements that have an impact on the carbon footprint (Table \ref{tab:carbon_footprint}).

For the construction phase, we will use the total mass of the optics as an indicator. The greater the glass mass, the greater the carbon footprint (related to extraction and polishing).

For the operation phase, we will use the total number of binned pixels as an indicator. The more pixels there are, the more infrastructure will be needed to store and process the acquired data.

\begin{table}[H]
    \centering
    \begin{tabular}{l|c|c|c|c|c}
         & FSS & 3x9 & 4x6 & 4x9 & 4x9Y \\
    \hline
    $N_{spectro}$ & 41 & 48 & 58 & 58 & 53 \\
    $M_{optics}^{unit}$ (kg) & 420 & 321 & 142 & 183 & 160 \\
    $M_{optics}^{total}$ (t) & 17.2 & 15.4 & 8.2 & 10.6 & 8.5 \\
    \hline\hline
    $N_{pix}^{total}$ (Gpx) & 1.97 & 5.19 & 3.71 & 8.35 & 7.63 \\
    $N_{pix, binned}^{total}$ (Gpx) & 0.78 & 1.26 & 1.01 & 1.52 & 1.51
    \end{tabular}
    \caption{Summary of the Environmental Impact drivers.}
    \label{tab:carbon_footprint}
\end{table}

\subsection{Method, weights, and results}

The scoring system ranges from 1 to 5. The higher the score, the more favourable it is to retain it. As an example, a high score in cost actually reflects a low cost.

Each quantity of interest is therefore scored. If there are several quantities of interest, these will be averaged with no weighting.

\begin{equation}
    \text{Score}_{\text{to maximise}}  = 1 + 4 \frac{x-\text{Worst}}{\text{Best}-\text{Worst}}
    \label{x_max}
\end{equation}

\begin{equation}
    \text{Score}_{\text{to minimise}}  = 1 + 4 \frac{\text{Worst} - x}{\text{Worst} - \text{Best}}
    \label{x_min}
\end{equation}

We distinguish two cases: when we want to maximise a quantity $x$, we use Equation \ref{x_max}. And when we want to minimise it, we use Equation \ref{x_min}.

\begin{figure}[h]
    \centering
    \includegraphics[width=0.9\linewidth]{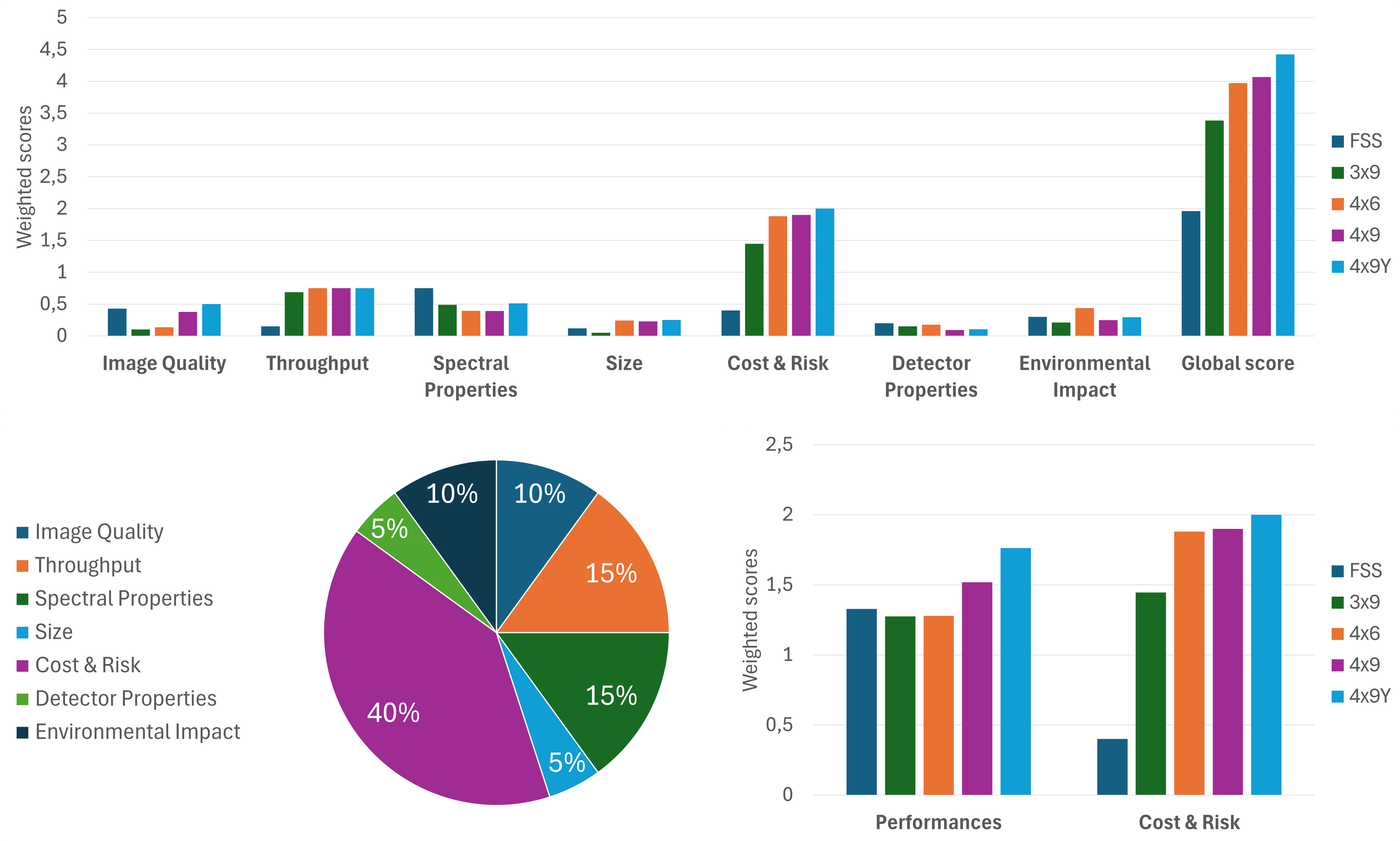}
    \caption{At the top, the weighted scores for the trade-off. At the bottom left, the trade-off matrix. And at the bottom right, a comparison between the performance scores and cost \& risk.}
    \label{fig:results}
\end{figure}

To obtain the final score, we this time take a weighted average from the scores obtained in each topic. The weighted scores, the distribution of weights, and the comparison between Performances and Cost \& Risk are presented in Figure \ref{fig:results}.

The 4x9Y is the design with the best score. Then, we choose to take the 4x9Y as the reference design and we keep the 4x9 as the backup design.

\section{Mechanical implementation}
\label{mecha}

\subsection{Rack arrangement}

It is simpler to integrate and align the optics on a bench. But we saw in Section \ref{sec:tradeoff} that the useful surface area would be far too large to accommodate all the modules on the azimuthal platform. For this reason, we are studying a rack arrangement.

\begin{figure}[h]
    \centering
    \includegraphics[width=0.6\linewidth]{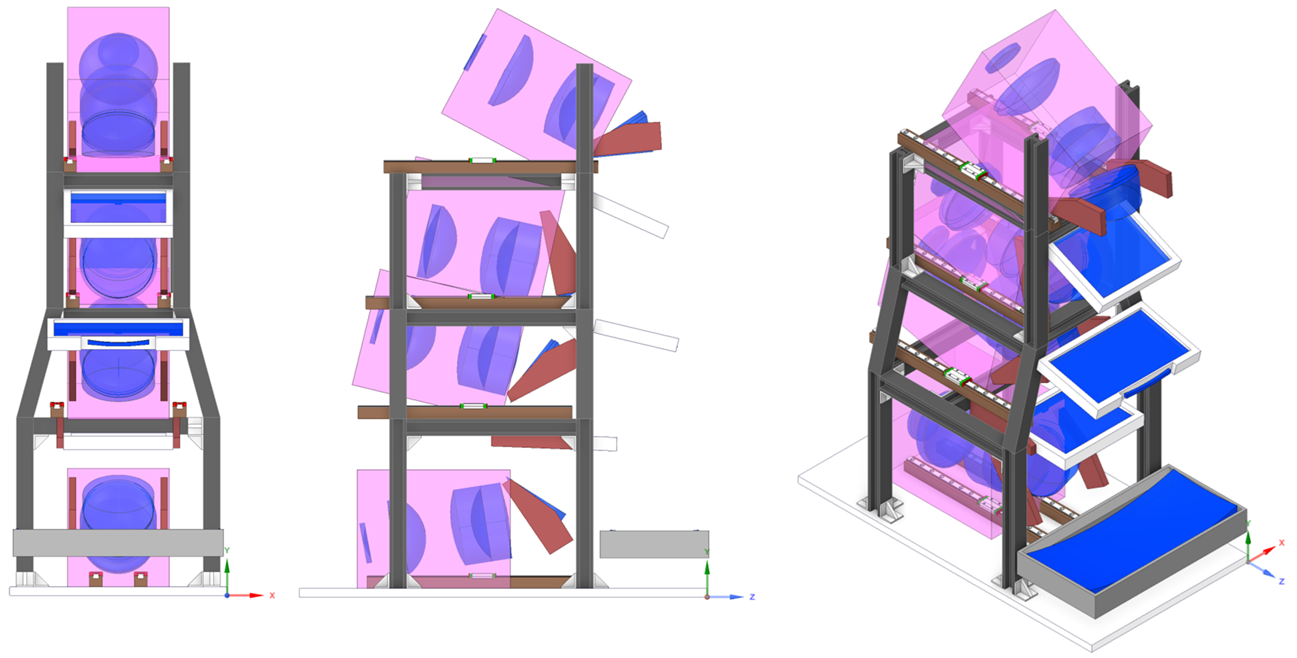}
    \caption{Example of a rack integration for the 4x6.}
    \label{fig:chassis}
\end{figure}

Here a rack does not refer to the arrangement of DESI where there are two levels of spectrographs. Rather to a chassis with several slots where the already assembled and aligned cameras could be slid in. A version of this chassis was produced for the first version of the 4x6 (at that point, the IZ arm had only one doublet and a lens) as shown in Figure \ref{fig:chassis}.

\begin{figure}[h]
    \centering
    \includegraphics[width=0.7\linewidth]{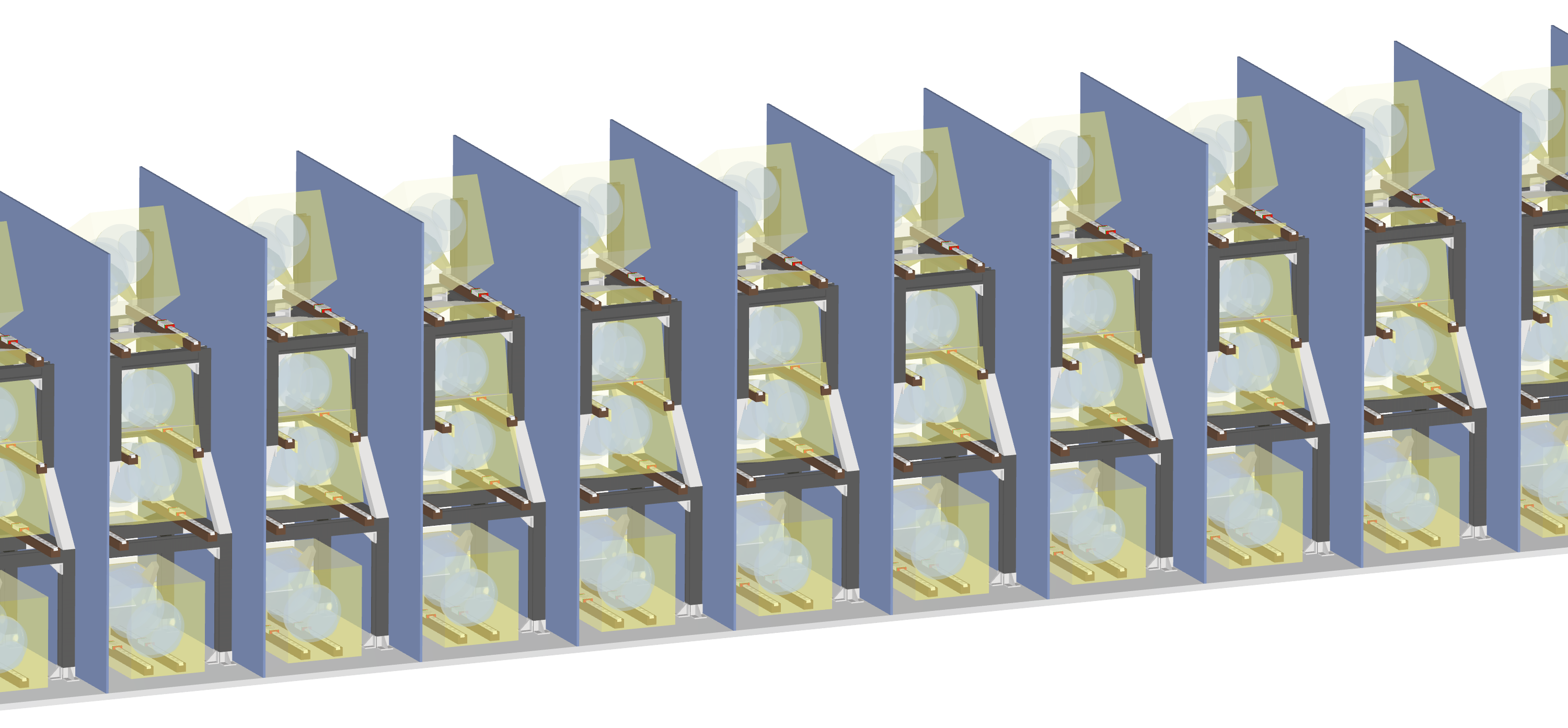}
    \caption{Chain assembly of the 4x6 racks.}
    \label{fig:rack_chain}
\end{figure}

\begin{figure}[h]
    \centering
    \begin{tabular}{cc}
         \includegraphics[width=0.35\linewidth]{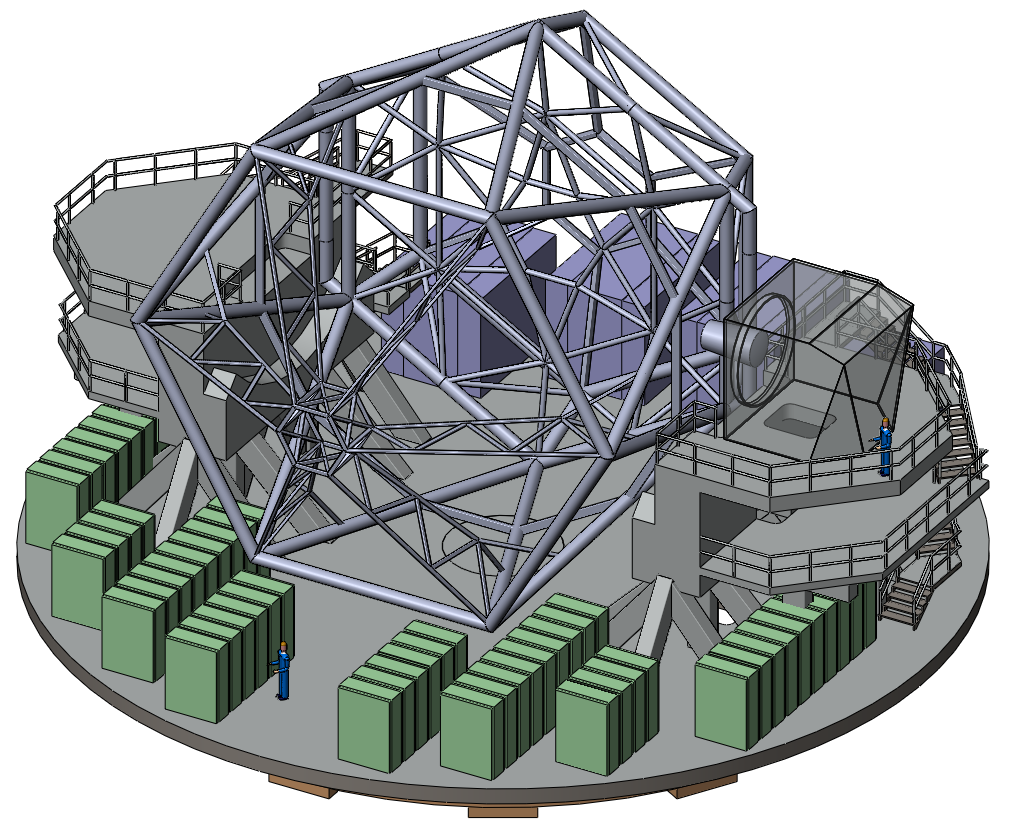} &
         \includegraphics[width=0.35\linewidth]{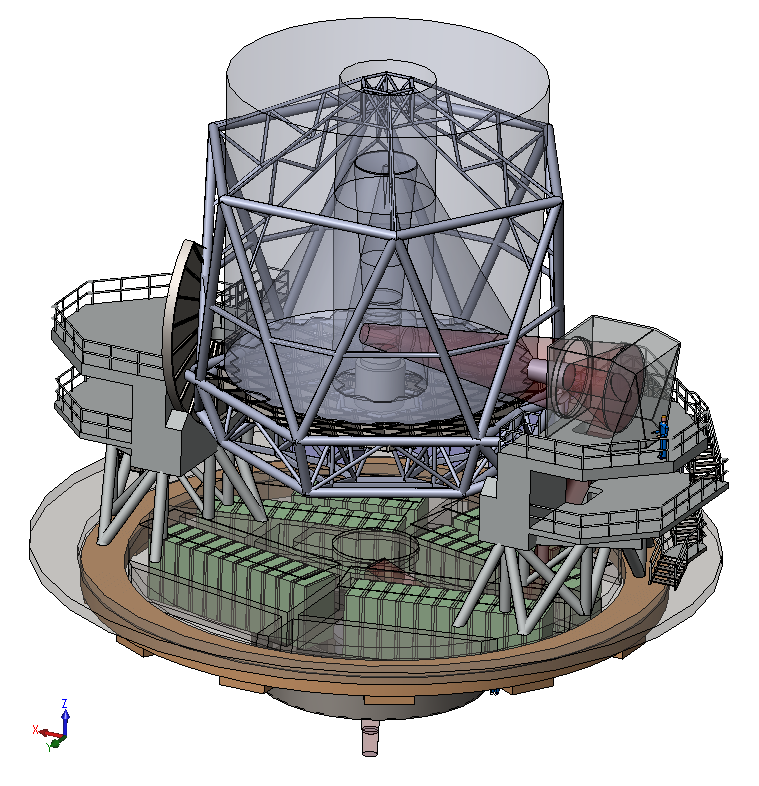}
    \end{tabular}
    \caption{Arrangement around the telescope [\citenum{Gausach2026}]. Left: on the azimuthal platform. Right: beneath the telescope in a MOS-LR room.}
    \label{fig:rack_tel}
\end{figure}

The barrel cameras are shown in pink in the Figure \ref{fig:chassis} and in yellow in the Figure \ref{fig:rack_chain}. From this design, we can already begin to work on more advanced issues such as the series integration of modules (Figure \ref{fig:rack_chain}). Or their arrangement in the observatory (Figure \ref{fig:rack_tel}), in particular whether they are placed around the telescope or in a MOS-LR room beneath the telescope. Several options are currently being studied for the racks. However, we must keep as our objective a simple system facilitating our access to the spectrograph components, such as the slit, the detectors, or the cryostats for maintenance. We must also minimise the fibre length, which has a loss of 1\% per metre.

\subsection{Bench arrangement}

\begin{figure}[H]
    \centering
    \includegraphics[width=0.8\linewidth]{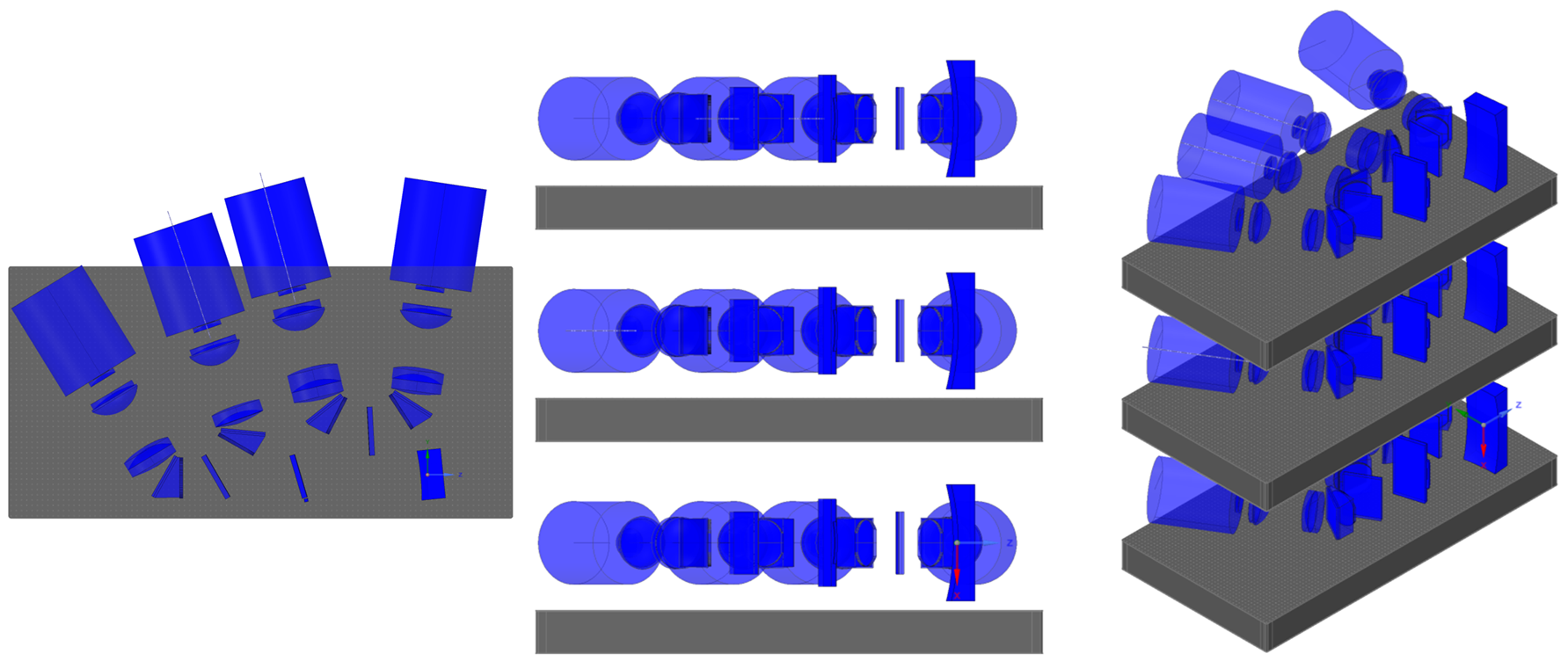}
    \caption{Example of a bench integration for the 4x9Y.}
    \label{fig:bench}
\end{figure}

One of the problems with a rack arrangement is gravity, which complicates the opto-mechanical part of the spectrographs. A more conservative solution could be studied using benches. In the manner of DESI, we could stack several benches (Figure \ref{fig:bench}). Looking at Table \ref{tab:V_S}, we realise that on average three benches would need to be stacked to cover the same area as three racks placed side by side. However, stacking a large number of spectrographs can also pose a vibration problem. Which can surely be resolved if the spectrographs are placed in a MOS-LR room beneath the telescope. This room would be equipped with a structure fixed from floor to ceiling onto which the benches could be slid like drawers. Solving the problems of access, maintenance, integration, and feasibility. This proposal has not yet been explored in detail but could constitute a backup if the rack arrangement presents too much risk. Furthermore, MUST appears to have chosen this option [\citenum{Cai2025}]. 

\section{Current reference design}
\label{sec:evol}

\begin{figure}[h]
    \centering
    \begin{tabular}{c}
         \includegraphics[width=0.6\linewidth]{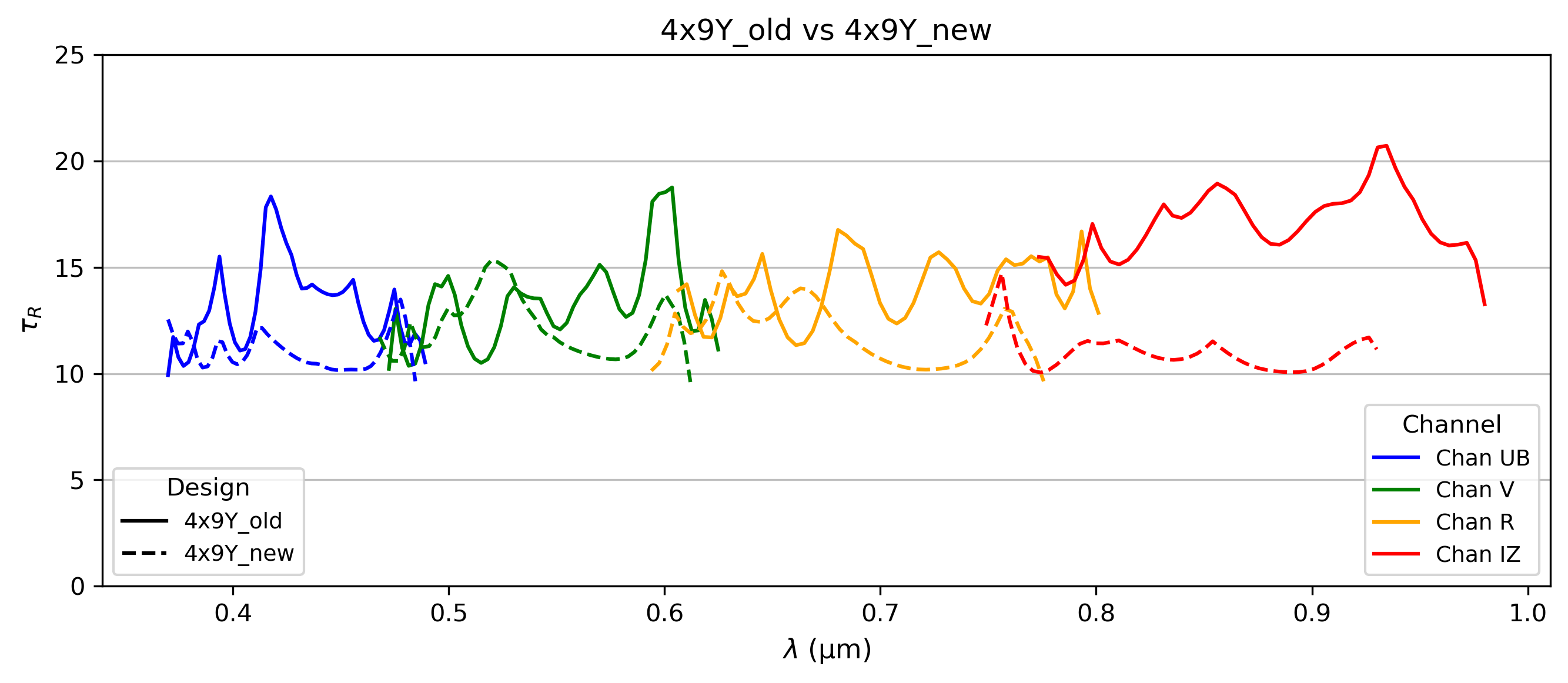} \\
         \includegraphics[width=0.6\linewidth]{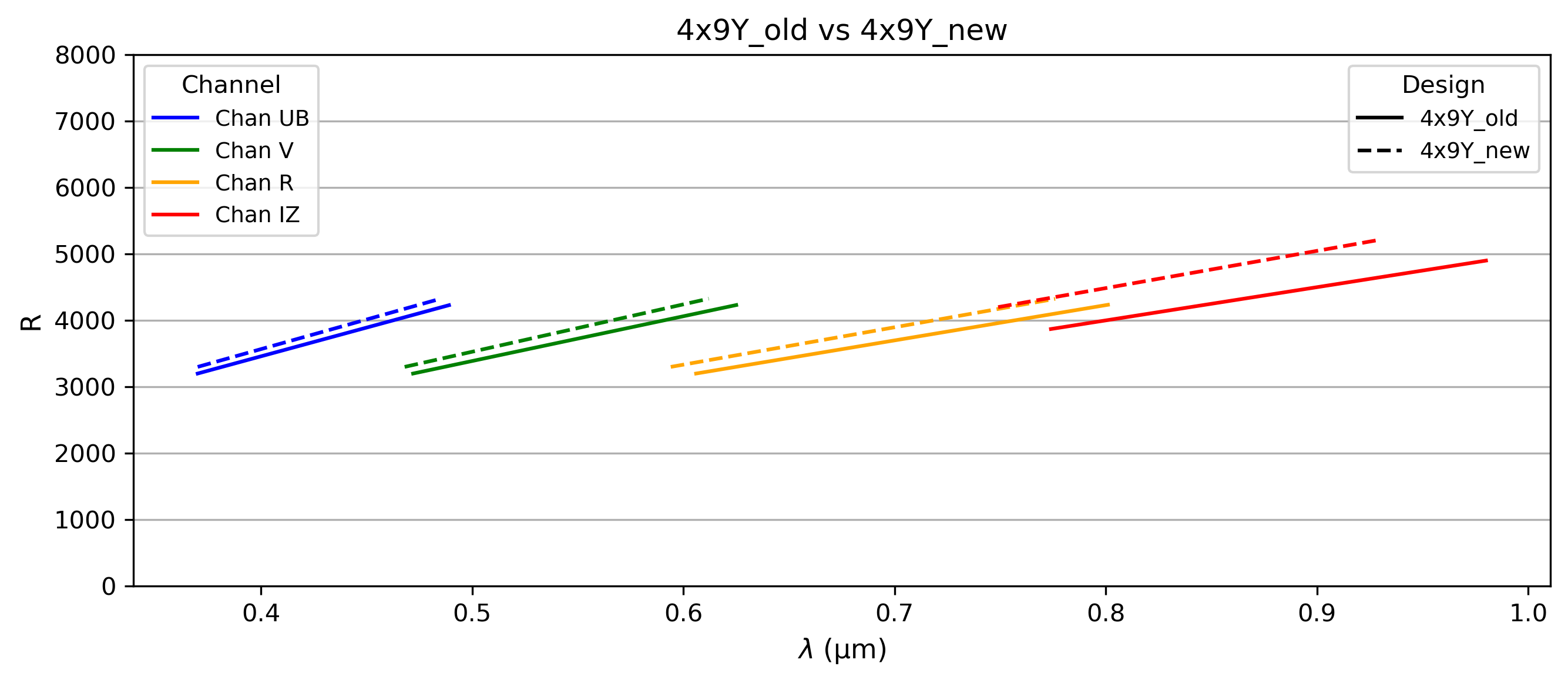}
    \end{tabular}
    \caption{Impact of the change of requirements and the detector clearance constraint on the image quality (left) and spectral resolution (right) of the 4x9Y.}
    \label{fig:UpgradeYAG}
\end{figure}

As outlined in Section \ref{sec:requir}, the spectral range was revised to 370–930 nm due to detector limitations. Then, we gain in spectral resolution but lose in image quality (Figure \ref{fig:UpgradeYAG}). In reality, this loss in image quality is not due to the change of requirements, but due to a constraint on the detector clearance (Section \ref{sec:design}). Previously, all our designs had a detector clearance of 1 mm. Which is too small for the electronics and the cryostats. We therefore increased this distance to 3.5 mm, which degraded the image quality.

This degradation in image quality and the increase in detector clearance is also accompanied by stronger asphericity in the optical design. This asphericity allows us to remain within the requirements at the expense of a higher cost and manufacturing risk.

Finally, these changes make the cameras slightly slower, going from f/1.6 to f/1.63. Which has the consequence of increasing the number of spectrographs to 55 (previously 53).

\begin{figure}[H]
    \centering
    \includegraphics[width=0.8\linewidth]{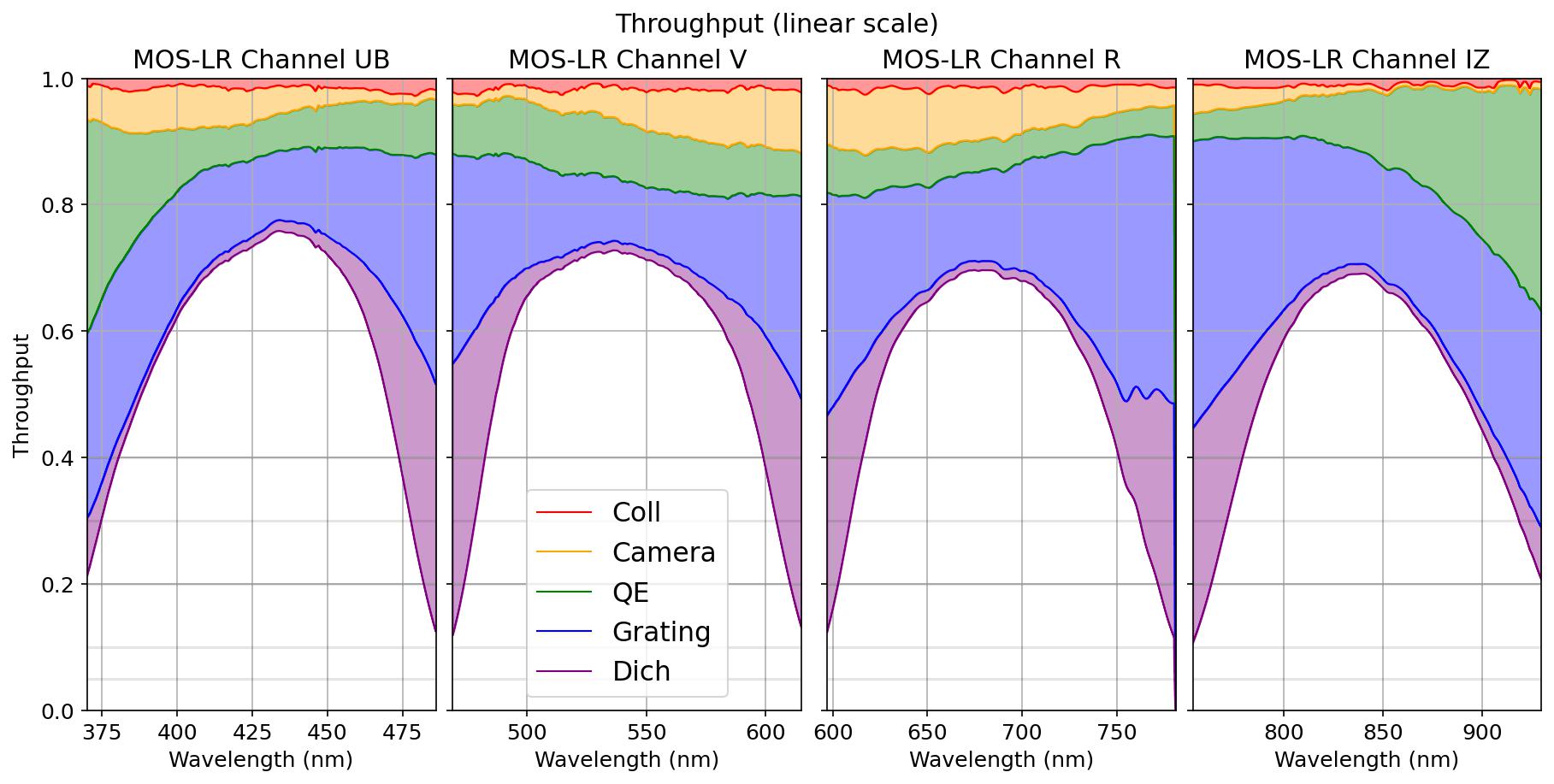}
    \caption{Throughput detail for each camera of the latest 4x9Y.}
    \label{fig:details_tp}
\end{figure}

More in-depth work on the MOS-LR transmission has been carried out, using data provided by institutes rather than mathematical models. Figure \ref{fig:details_tp} also takes stock of the transmission of each individual optical element.

\section{CONCLUSION}

Our trade-off analysis shows that slower cameras are easier to manufacture due to lower risk and smaller lens diameters. Beyond this, the use of YAG as an innovative material for the field lenses allows us to improve image quality without significantly increasing the size and speed of the cameras. However, with slower cameras we require a larger number of spectrographs, raising challenges regarding their integration and distribution, the two aspects being closely linked. These aspects are currently under investigation as part of the ongoing pre-study.

\acknowledgments 
 
This project has been funded by the European Union’s Horizon Europe research and innovation programme under grant agreement No. 101183153. 

\bibliography{report} 
\bibliographystyle{spiebib} 

\end{document}